\documentclass[fleqn,usenatbib]{mnras}

\usepackage{newtxtext,newtxmath}

\usepackage[T1]{fontenc}

\DeclareRobustCommand{\VAN}[3]{#2}
\let\VANthebibliography\thebibliography
\def\thebibliography{\DeclareRobustCommand{\VAN}[3]{##3}\VANthebibliography}

\usepackage{graphicx}	
\usepackage{amsmath}	
\usepackage{amssymb}	

\title[Detections of NHD and ND$_2$]{Deuterium fractionation of nitrogen hydrides: detections of NHD and ND$_2$}

\author[A. Bacmann et al.]{
A. Bacmann,$^{1}$\thanks{E-mail: aurore.bacmann@univ-grenoble-alpes.fr}
A. Faure,$^{1}$
P. Hily-Blant,$^{1}$
K. Kobayashi,$^{2}$
H. Ozeki,$^{3}$
S. Yamamoto,$^{4}$
L. Pagani$^{5}$
and F. Lique$^{6}$
\\
$^{1}$Univ. Grenoble Alpes, CNRS, IPAG, 38000 Grenoble, France\\
$^{2}$Department of Physics, University of Toyama, 3190 Gofuku, Toyama 930-8555, Japan\\
$^{3}$Department of Environmental Science, Faculty of Science, Toho University, 2-2-1 Miyama, Funabashi, Chiba 274-8510, Japan \\
$^{4}$Department of Physics, The University of Tokyo, 7-3-1, Hongo, Bunkyo-ku, Tokyo 113-0033, Japan\\
$^{5}$LERMA \& UMR8112 du CNRS, Observatoire de Paris, PSL University, Sorbonne Universit\'s,  CNRS, F-75014 Paris, France\\
$^{6}$LOMC - UMR 6294, CNRS-Universit\'e du Havre, 25 rue Philippe Lebon, BP 1123, F-76063 Le Havre Cedex, France
}
\date{Accepted 2020 September 11. Received 2020 September 11; in original form 2020 July 10}

\pubyear{2020}

\begin{document}
\label{firstpage}
\pagerange{\pageref{firstpage}--\pageref{lastpage}}
\maketitle

\begin{abstract}
Although ammonia is an abundant molecule commonly observed towards the dense
   interstellar medium, it has not yet been established whether
   its main formation route is from gas-phase ion-molecule reactions or grain-surface hydrogen
   additions on adsorbed nitrogen atoms. Deuterium fractionation can be used as a tool  to 
   constrain formation mechanisms. High abundances of deuterated molecules are routinely observed
   in the dense interstellar medium, with the ratio between deuterated molecules and the main isotopologue enhanced
   by several orders of magnitude with respect to the elemental
   D/H ratio. In the case of ammonia, the detection of its triply deuterated
   isotopologue hints at high abundances of the deuterated  intermediate nitrogen radicals, 
   ND, NHD and ND$_2$. So far however, only ND has been detected in the interstellar medium.  In this paper, to constrain the formation of ammonia, we aim at determining the  NHD/NH$_2$ and ND$_2$/NHD abundance ratios,  and compare them with the predictions of both pure gas-phase
   and grain-surface chemical models. We searched
   for the fundamental rotational transitions of NHD and ND$_2$ towards the
   class 0 protostar IRAS16293-2422, towards which NH, NH$_2$ and ND
   had been previously detected. Both  NHD and ND$_2$ are detected in absorption towards the source. The
   relative abundance ratios NH$_2$ : NHD : ND$_2$ are close to  8 : 4 : 1. These
   ratios can be reproduced by our gas-phase chemical model within a
   factor of two-three. Statistical ratios as expected from grain-surface chemistry are 
   also consistent with our  data.  Further investigations of the ortho-to-para ratio in ND$_2$ , both theoretical and observational, could bring new constraints to better understand nitrogen hydride chemistry.
\end{abstract}




\section{Introduction}

High degrees of deuterium fractionation in interstellar molecules have long been observed towards  star forming regions, in particular towards the cold and dense prestellar cores and envelopes of protostars. Indeed, because of the lower zero-point energy of H$_2$D$^+$ compared to that of H$_3^+$, the  deuterium exchange reaction  H$_3^+$ + HD $\rightarrow$ H$_2$ + H$_2$D$^+$ is slightly exothermic and favoured at temperatures lower than 20\,K, increasing the H$_2$D$^+$/H$_3^+$ ratio and therefore the possibility of transfering a deuterium atom to molecular species by ion-neutral chemistry. This ratio is further enhanced when CO, which is a major destroyer of the 
H$_3^+$ ions, is frozen out on dust grains and undergoes a drastic abundance drop. In this case, the above-mentioned reaction becomes the  main reaction destroying H$_3^+$, which 
leads to an increased H$_2$D$^+$/H$_3^+$  ratio, reaching  sometimes unity,  according to some models \citep{1992A&A...258..479P,2003ApJ...591L..41R}. Similar reactions with all deuterated isotopologues of H$_3^+$ also take place, leading to high abundances of H$_2$D$^+$, D$_2$H$^+$, and D$_3^+$ \citep{2003ApJ...591L..41R,2004A&A...418.1035W}. 

Deuterium fractionation can theoretically be used to constrain molecular formation pathways, because the ratios between the deuterated species and the main isotopologue are expected to be different if the molecule forms in the gas phase or as a result of grain surface chemistry. In the latter case, deuteration is expected to follow a statistical scheme, as highlighted in, e.g, \citet{1989MNRAS.240P..25B}. The deuteration of such a ubiquitous molecule as ammonia has stimulated many studies, especially since the discovery of its triply deuterated isotopologue by \citet{2002ApJ...571L..55L} and \citet{2002A&A...388L..53V} with an abundance ratio ND$_3$/NH$_3\sim0.001$, an increase of 12 orders of magnitude with respect to the elemental D/H ratio \citep{2006ApJ...647.1106L}, challenging chemical models. As likely precursors of ammonia, nitrogen hydrides can bring valuable clues to constrain ammonia formation mechanisms. \citet{2015A&A...576A..99R} argue that highly-deuterated ammonia can form from pure gas phase chemistry. Observations of deuterated nitrogen hydride radicals like NHD and ND$_2$ can help to test this type of scenario.

The {\it Herschel Space observatory}\footnote{{\it Herschel} is an ESA space observatory with science instruments provided by European-led Principal Investigator consortia and with important participation from NASA} has allowed us to access the fundamental rotational transitions of nitrogen hydrides. Since its launch in 2009, both the NH radical and its deuterated counterpart have been detected towards the low-mass class 0 protostar IRAS16293-2422 \citep{2010A&A...521L..42B} and towards the dark core IRAS16293E \citep{2016A&A...587A..26B}. The abundances derived for these species have been found to be consistent with the model predictions of \citet{2005A&A...438..585R} and \citet{2015A&A...576A..99R}. The NH$_2$ radical has also been detected towards IRAS16293-2422 \citep{2010A&A...521L..52H} but no detection of its singly isotopologue NHD or doubly deuterated isotopologue ND$_2$  
had been reported until recently.  However, while this manuscript was under revision, \citet{2020arXiv200707504M}  presented  the detection of the two latter species in IRAS16293-2422. 

In this paper, we report the detection of both NHD and {\it para-}ND$_2$ towards IRAS16293-2422, based on different transitions from those in \citet{2020arXiv200707504M}, and discuss the implications of the derived abundances and abundance ratios on the formation of nitrogen hydrides. The observations are presented in Section\,\ref{obs}. In Section\,\ref{analysis}, we derive the abundances and discuss them in the light of new chemical models in Section \,\ref{mod}, before concluding in Section\,\ref{conclu}.

\section{Observations}
\label{obs}

The search for NHD and ND$_2$ was conducted towards IRAS16293-2422 which is a class 0 low-mass protostar where NH, ND and NH$_2$ had already been discovered \citep{2010A&A...521L..42B,2010A&A...521L..52H}. The coordinates of integration for the observations are $\alpha_{2000}=16^h$32$^m$22.8$^s$ $\delta_{2000}=-24{\degr}28^\prime33^{\prime\prime}$, the same as those where the  nitrogen hydrides were previously detected. While the source is a binary (composed of two sources A and B) \citep[see][for an extensive description of the source]{2016A&A...595A.117J}, the separation between the components is 5.1$\arcsec$, so that our observations encompass both components. The species NH, ND and NH$_2$ were all seen in absorption against a background emitted by the warm dust. This is partly due to the very high critical densities of the fundamental rotational transitions of nitrogen hydrides, especially NH$_2$ and NH \citep[around 10$^{7}$\,cm$^{-3}$, see][for the derivation of the collisional rate coefficients]{2012JChPh.137k4306D,2019MNRAS.tmp.2223B}, and the presence of the species in the lower density envelope. Indeed, the frequency of the fundamental rotational transition of ND is a factor of 2 smaller than for NH, and therefore the critical density a factor of $\sim 10$ smaller, making ND easier to be seen in emission than its hydrogenated counterpart. 

\subsection{Spectroscopy}

Microwave spectroscopy measurements and analysis of NHD  have been performed by \citet{1997JChPh.107.9289K} and \citet{2013mss..confETH01M}, while those for ND$_2$ have been performed by \citet{1991JChPh..94.3423K} and \citet{2017ApJS..233...15M}. The spectra of both radicals have a complex fine and  hyperfine structure due to the interaction with the electronic and nuclear spins of nitrogen, deuterium and hydrogen (the latter only for NHD). For NHD, we targetted the group of hyperfine components ($N_{K_\mathrm{a} K_\mathrm{c}} J: 1_{0 1}\, 3/2 - 0_{0 0}\, 1/2$)\ around 412.7\,GHz. The $N_{K_\mathrm{a} K_\mathrm{c}} J: 1_{0 1}\, 1/2 - 0_{0 0}\, 1/2$ transition at 413.5\,GHz was additionally present in our observing bandwidth. For ND$_2$, the transitions ($N_{K_\mathrm{a} K_\mathrm{c}} J: 1_{1 1}\, 1/2 - 0_{0 0}\, 1/2$) at 531\,GHz were observed, which correspond to {\it para}-ND$_2$. The frequencies of the transitions and their Einstein coefficients for spontaneous emission are given in Table\,\ref{spectro_nhd1} and \ref{spectro_nhd2} for NHD and in Table\,\ref{spectro_nd2} for ND$_2$.

\subsection{NHD observations}

Singly deuterated amidogen was searched for with the APEX telescope\footnote{This publication is based on data acquired with the Atacama Pathfinder Experiment (APEX). APEX is a collaboration between the Max-Planck-Institut fur Radioastronomie, the European Southern Observatory, and the Onsala Space Observatory.} 
on Chajnantor Plateau, Chile.
The observations were carried out in April 2013 in service mode with the Swedish heterodyne receiver instrument 
APEX-3 \citep{2008A&A...490.1157V} tuned to 413.1\,GHz in the upper sideband. The receiver was connected to the XFFTS spectrometer
covering the entire receiver band with a spectral resolution of 0.076 kHz, corresponding to a velocity resolution of
0.055\,km\;s$^{-1}$ at this frequency.  

During the observing runs, the amount of precipitable
water vapor was mostly between 0.8\,mm and 1.2\,mm, resulting in system temperatures between 400 and 500\,K
depending on source elevation. Because the NHD signal is expected to be moderately extended, the position 
switching mode was used with an offset of 70\arcsec\ in Right Ascension to the West of the source, in a direction where the density
of the protostellar envelope and the molecular emission drops quickly \citep{2001A&A...375...40C}. The data were reduced
using the Gildas/CLASS software\footnote{http://www.iram.fr/IRAMFR/GILDAS}: the spectra were averaged together and a low
order polynomial was fitted to line-free regions of the spectra and subtracted.
The forward efficiency of the antenna was taken to be 0.95 and the beam efficiency 0.65, an intermediate value between
the measured value at 352\,GHz and that measured at 464\,GHz \citep{2006A&A...454L..13G}. These efficiencies were
used to convert the spectra from the $T_a^*$ to the $T_{\rm mb}$ scale. The APEX beam size is 15\arcsec\ full width at 
half maximum (FWHM) at the observing frequency. The {\sl rms} noise in the spectrum is $\sim$\;33\,mK ($T_{\rm mb}$ scale)
in 0.11\,km\,s$^{-1}$ channels.

\subsection{ND$_2$ observations}
The search for doubly deuterated amidogen was carried out with the {\it Herschel}  Space Observatory \citep{2010A&A...518L...1P} in
the course of an open time project (PI: P. Hily-Blant, OBSID 1342227404). The ND$_2$ $N_{K_\mathrm{a} K_\mathrm{c}} J: 1_{1 1}\, 1/2 - 0_{0 0}\, 1/2$ line at 531\,GHz was observed with the HIFI instrument \citep{2010A&A...518L...6D} in band 1a 
on 26 August 2011 (OBSIDs 1342227403 and 1342227404), with both the wide band spectrometer (WBS), and the high resolution spectrometer (HRS). The WBS has a spectral resolution of 1.1\,MHz, corresponding to a velocity resolution of 0.6\,km\,s$^{-1}$ at the frequency of the observations, and the HRS has a resolution of 0.25\,MHz, corresponding
to a velocity resolution of 0.14\,km\,s$^{-1}$. The observations were performed in the single pointing dual beam switch mode with fast chopping (4\,Hz)
and optimization of the continuum. This improves the subtraction of the standing waves and the determination of the continuum. In
this mode, the OFF positions are at fixed offsets 3$\arcmin$ away from the source coordinates on either side (East-West) of the source. We checked on the spectra
of these OFF positions that no ND$_2$ signal was present. The data were processed with the HIPE pipeline \citep{2010ASPC..434..139O} version 14 up to level\,2.5 products, after which they were exported to Gildas/CLASS data format for further analysis: the spectra in both horizontal and vertical polarizations were averaged, and a first order polynomial baseline  was fitted to line-free regions of the spectrum and subtracted.  The HIFI beam at 531\,GHz, the frequency of the ND$_2$ transitions, is $\sim 39$\arcsec\ FWHM, the main beam efficiency $\eta_\mathrm{mb}$ is 0.62 and the forward efficiency $\eta_\mathrm{fwd}$ is 0.96  \citep{2017A&A...608A..49S}. The {\sl rms} noise in the spectrum is $\sim$\;6.6\,mK ($T_{\rm mb}$ scale) in 0.56\,km\,s$^{-1}$ channels.

\subsection{Continuum data}

In this study, we also used continuum data (see section\,\ref{continuum}).  Maps at 250\,$\mu$m, 350\,$\mu$m, and 500\,$\mu$m 
observed by  {\it Herschel}/SPIRE \citep{2010A&A...518L...3G} were taken from the {\it Herschel} Gould Belt survey  \citep[HGBS][]{2010A&A...518L.102A}.
We used the level 2.5 maps of the L1689 cloud (OBSID 1342239773)  processed by the HIPE pipeline (version 12.0) and calibrated for extended sources. Absolute calibration  using HFI data from the Planck satellite is also performed by the pipeline. We also subtracted from the images a constant offset measured in an apparently emission-free region to the South-East of the L1689 cloud. This serves as a first-order removal of the emission from the diffuse galactic background.  Because the {\it Herschel}/SPIRE maps focus on a $\sim 15\arcmin\times 15$\arcmin region centered on IRAS16293-2422 and do not include areas outside of the Rho Ophiuchi cloud, we used the larger-scale maps of Rho Ophiuchi covering the whole complex (OBSID 1342205093 and 1342205094, also part of the HGBS) to measure the off-cloud emission for each of the three wavelengths. These values were corrected from the intensity offsets between the small maps of IRAS16293-2422 and the large-scale maps of Rho Ophiuchi by comparing the map intensities in a region common to both types of maps. 

The beam sizes (FWHM) at wavelengths 250, 350, and 500\,$\mu$m are 18\arcsec, 25\arcsec and 36\arcsec, respectively. As in \citet{2016A&A...587A..26B}, additional ground-based maps of the region at 850\,$\mu$m taken with the  SCUBA-2 bolometer array on the {\it James Clark Maxwell telescope} (JCMT) as part of the JCMT Gould Belt Survey \citep{2013MNRAS.430.2513H}  were used. The data and the data reduction procedure are described in \citet{2015MNRAS.450.1094P}. Finally, we also used the map at 1.2\,mm taken by  \citet{2002ApJ...569..322L} with the MAMBO-II bolometer on IRAM\,30m telescope. The FWHM beamsizes are $\sim 14$\arcsec\ for the JCMT/SCUBA-2 map and $\sim 11$\arcsec\ for the IRAM 30\,m data.

\section{Results and analysis}
\label{analysis}

\subsection{Continuum estimation\label{continuum}}

\subsubsection{NHD}

\begin{figure}
  \centering
  \includegraphics[width=8cm]{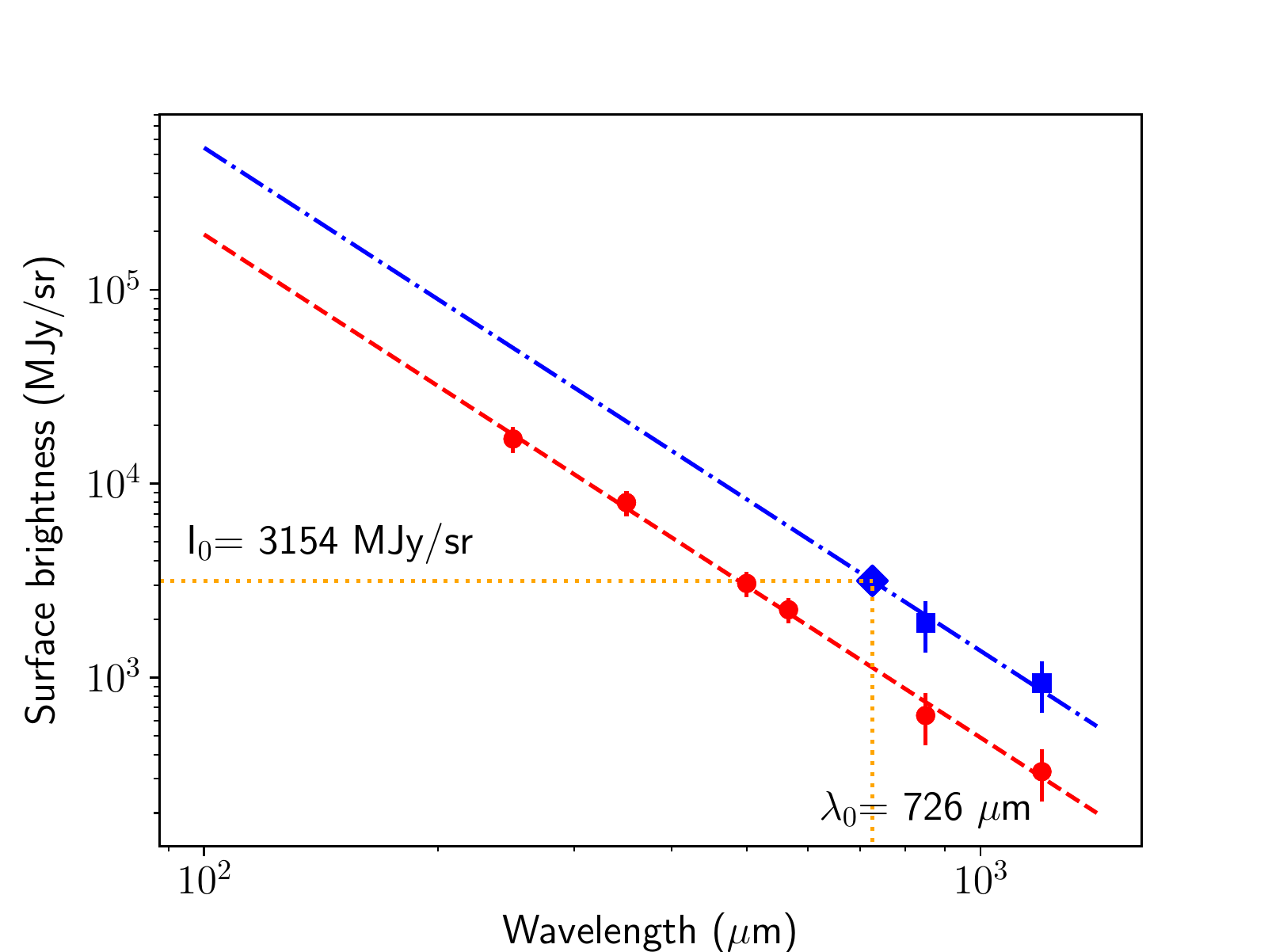}
  \caption{Continuum intensities of IRAS16293-2422 measured at 36$\arcsec$ resolution (red dots) and power-law fit (red dashed line). The index of the power-law is found to be $\alpha=-2.6$. The blue squares are the intensities at 
  15$\arcsec$ resolution, and the blue dash-dotted line fits these points with the same power-law index $\alpha$. The derived intensity $I_0$  at the wavelength of the APEX observations ($\lambda_0=726\,\mu$m) and at a resolution of 15$\arcsec$ is marked as a blue diamond.}
             \label{sed}%
  \end{figure}

Because the ground-based observations do not enable us to measure the continuum level of the NHD spectrum, we use available 
continuum observations taken at various wavelengths to infer the value at the frequency of the NHD transition (413\,GHz, which corresponds
to a wavelength of 726\,$\mu$m) by interpolation.  At wavelengths larger than typically 200\,$\mu$m, the spectral energy distribution of the protostar is close to the Rayleigh-Jeans regime, so that fitting a simple power-law to the source fluxes as a function of wavelength allows for a straightforward determination of the continuum value
at 413\,GHz. First the continuum maps were smoothed to a common angular resolution, i.e. 36$\arcsec$,  that of the coarsest map from SPIRE at 500$\mu$m. To achieve this, the 250\,$\mu$m, 350\,$\mu$m and 500\,$\mu$m SPIRE maps were convolved with the appropriate kernels provided by \citet{2011PASP..123.1218A}. The 850$\mu$m SCUBA-2 map and the 1.2\,mm MAMBO-II map were convolved with gaussian kernels of FWHM 33\,$\arcsec$ and 34.3$\arcsec$, respectively, the former also from \citet{2011PASP..123.1218A}. The average surface brightnesses of the source at each wavelength were measured in an aperture of 15$\arcsec$, corresponding to the beam of the NHD observations. We also added the continuum intensity from the {\it Herschel}/HIFI ND$_2$ spectrum at 531\,GHz (565$\mu$m), which is at a similar resolution. A power-law of index $\alpha=-2.6$ could be fitted to the data points (Fig.\,\ref{sed}), which enabled us to interpolate the intensity at 413\,GHz at a resolution of 36$\arcsec$. To derive the source intensity at the resolution of the APEX observations (15$\arcsec$), we smoothed the SCUBA-2 map to the resolution of the APEX observations by convolution with a gaussian kernel of 5$\arcsec$ from \citet{2011PASP..123.1218A}, and the MAMBO-II bolometer map by convolution with a gaussian kernel of 10.2$\arcsec$ FWHM. Those two points were fitted with a power-law with the same exponent $\alpha$, and from this power-law, the value of the surface brightness was obtained at 413\,GHz at a resolution of 15$\arcsec$. This method assumes that the dust opacity index does not depend on the angular resolution, i.e. on the dust temperature or density in the ranges probed by the observations. While this might not be true in the general case, this approximation is good enough for our purpose here, considering the uncertainties in the data analysis. The interpolated surface brightness is 3154\,MJy/sr, corresponding to a brightness temperature $T_c (\mathrm{NHD})=600$\,mK. We used the latter value as the continuum brightness temperature of the source in the APEX spectrum.

Because of the observing technique used to remove the strong atmospheric emission for ground-based observations, the emission arising from extended structures in the astrophysical source is partly filtered out.  Since we are interested in the mean intensity of the protostar in a compact region around its maximum, our surface brightness values at 850\,$\mu$m and 1.2\,mm should not be too affected. Nevertheless, the filtering of the extended emission for the JCMT and IRAM 30\,m maps could lead to an underestimate of the surface brightness at 850\,$\mu$m and 1.2\,mm. Therefore, the power-law that we fitted could in fact be shallower and the continuum intensity at 413\,GHz could be slightly overestimated.

\subsubsection{ND$_2$}

For ND$_2$, the continuum was measured by fitting a zeroth order polynomial to the spectrum  
in regions that do not show any spectral lines. Assuming that the observed continuum is the sum of that in the upper side-band and that in the lower side-band, and supposing that it increases with frequency following a power-law of index $\alpha=-2.6$, we find that the single side band continuum at 531\,GHz is $T_c (\mathrm{ND_2})=210$\,mK. 

\subsection{NHD and ND$_2$ spectra}

\begin{figure}
   \centering
   \includegraphics[width=8cm]{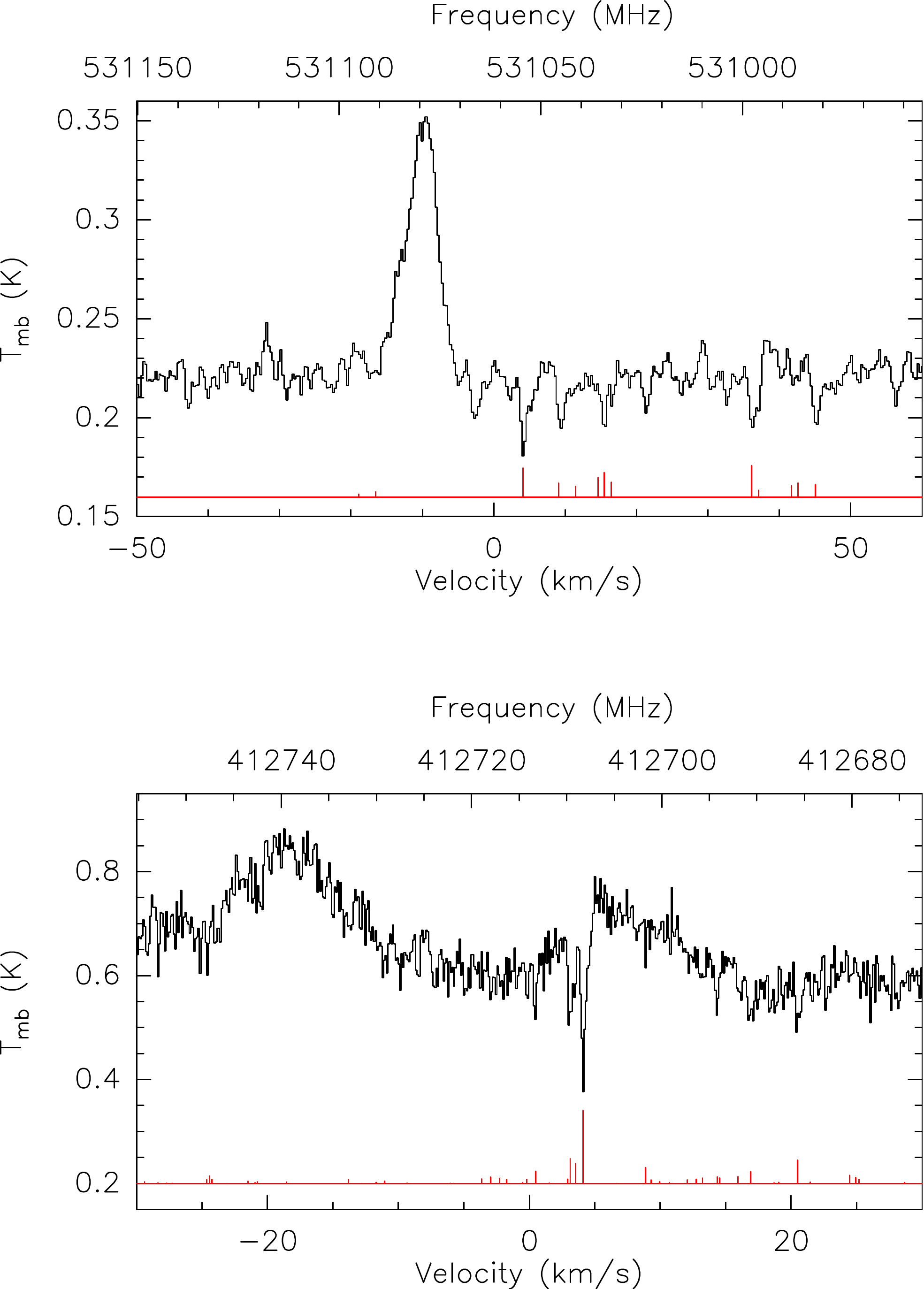}
   \caption{NHD (bottom panel) and ND$_2$ (top panel) spectra in IRAS16293-2422. The red sketch shows the positions and expected relative
   intensities of the hyperfine components. The strong emission feature at a velocity of $-$10\,km s$^{-1}$ (corresponding to a frequency of 531079.2\,MHz) 
   in the ND$_2$ spectrum is likely the $N_{K_\mathrm{a} K_\mathrm{c}}: 11_{1\;11}-10_{1\,10}$ transition of CH$_3$OH. The broad features in the NHD spectrum at velocities $-17$\,km\,s$^{-1}$ and 
   7\,km\,s$^{-1}$ are unidentified.}
              \label{nhdnd2}%
    \end{figure}

The observed NHD and para-ND$_2$ spectra are shown in Fig.\,\ref{nhdnd2}. For both spectra, the continuum levels derived in section\,\ref{continuum} were added to the spectra after baseline withdrawal. Both molecules are seen in absorption against the continuum. For the NHD spectrum, a rather broad emission feature (rest frequency 412702.5\,MHz) is visible at a similar velocity as that of the NHD lines. The FWHM of this feature is $\sim$\,8.5\,km s$^{-1}$, like for other similarly-looking features in the spectrum. Such wide spectral lines have already been seen  in this source \citep[e.g. TIMASSS,][]{2011A&A...532A..23C}, and are consistent with that of lines emitted in the hotter central region. However these broad emission lines unlikely arise from NHD itself: the near-gaussian feature against which the two strongest NHD features are seen in absorption peaks at a velocity $\sim$ 6.5\,km\,s$^{-1}$, higher than the hot corino velocity at 3.1\,km\,s$^{-1}$ and 2.7\,km\,s$^{-1}$ for sources A and B, respectively \citep{2011A&A...534A.100J,2016A&A...595A.117J}, and than the line velocities measured in the TIMASSS survey  (lower than 5\,km\,s$^{-1}$). In the following, we have considered that this feature arises from an unknown molecular transition (different from NHD), but we cannot fully exclude the possibility that it is a baseline ripple. The other strong, broad emission lines seen in Fig.\,\ref{nhdfit} (e.g., at  412738.7\,MHz, 413541.1\,MHz, and 413564.9\,MHz) are also unidentified and probably arise from the warm inner region.

Several hyperfine components are detected with the expected velocity shifts both for NHD ($J$: 3/2 $-$ 1/2) at 412.7\,GHz and para-ND$_2$ at 531\,GHz. No hyperfine component is seen for the NHD $J$: 1/2 $-$ 1/2 transition at 413.5\,GHz (Fig.\,\ref{nhdfit}). The source velocity is found to be 4.1\,km\;s$^{-1}$ for both NHD and ND$_2$. This is about 0.3\;km\;s$^{-1}$ higher than the velocities of 3.8\,km\;s$^{-1}$ for ND, NH and NH$_2$ in the same source \citep{2010A&A...521L..42B,2010A&A...521L..52H}, hinting at possibly different spatial locations for the deuterated forms of amidogen, or a lack of accuracy in the values of the rest frequencies. Because the noise in the spectra is relatively high, only the hyperfine components that are expected to be the strongest are detected, and no  theoretically weak component is detected, so that we are confident that both species are present. 

The fact that the lines are seen in absorption and their narrow linewidths ($\Delta v = 0.5$\,km\;s$^{-1}$ for NHD and $\Delta v\sim 0.7$\,km\;s$^{-1}$ for ND$_2$, though the latter value is subject to more uncertainty because of the lower signal-to-noise ratio) is consistent with the presence of both species in the low-density and low-temperature protostellar envelope, rather than in the higher density regions close to the hot core.

\subsection{Column density estimates}

Both NHD and ND$_2$ spectra were analysed in a similar fashion: the optical depth was derived as a function of frequency given a column density and assuming the same excitation temperature for all the hyperfine components. Because of the overlap between the hyperfine components, such a simple model may not be able to reproduce well the line relative ratios. More sophisticated and accurate treatments are not possible, since collisional coefficients are not yet available for these two species. The consequence on the column density determination should remain limited because of the low optical depths of the transitions. For Gaussian lines, the velocity integrated optical depth  is 
\begin{equation*}
\int{\tau(v) dv} = \tau_0 \Delta v \sqrt{\frac{\pi}{4\ln{(2)}}}
\end{equation*}
where $\tau_0$ is the line central optical depth and $\Delta v$ the line full width at half maximum.
The optical depth at line centre is therefore given by\\
\begin{equation*}
\tau_0 = N \frac{c^3}{8\pi\nu^3 }\frac{A_\mathrm{ul} g_\mathrm{up} }{Q(T_\mathrm{ex})}\frac{e^{h\nu/(k_\mathrm{b}T_\mathrm{ex})}-1}{e^{E_\mathrm{up}/k_\mathrm{b}T_\mathrm{ex}}}\frac{1}{\Delta v}\sqrt{\frac{4\ln{(2)}}{\pi}}
\end{equation*}
where $Q(T_\mathrm{ex})$ is the partition function at excitation temperature $T_\mathrm{ex}$, $N$ is the total molecular column density, $\nu$ the frequency of the transition, $A_\mathrm{ul}$ the Einstein spontaneous emission coefficient of the transition, $g_\mathrm{up}$ the upper level degeneracy, $c$ the speed of light, $E_\mathrm{up}$ the energy of the upper level, and $h$ and $k_\mathrm{b}$ are the Planck and Boltzmann constants, respectively. The total optical depth as a function of velocity is given by (neglecting the effect of line overlap on line excitation)\\
\begin{equation*}
\tau(v) = \sum_i \tau_0^i \exp{\left(-4\ln{(2)}\left(\frac{v-v_0}{\Delta v}\right)^2\right)}
\end{equation*}
where $v_0$ is the rest velocity, $i$ refers to hyperfine component $i$ and $\tau_0^i$ the optical depth of the hyperfine component $i$ is  $\tau_0^i=\tau_0 R_i$, with $R_i$ the hyperfine component statistical weight. 
The brightness temperature is then simply \\
\begin{equation*}
T_\mathrm{b}(v) = (J_\nu(T_\mathrm{ex})-J_\nu(T_\mathrm{CMB}))(1-e^{-\tau(v)})+ T_\mathrm{c}\ e^{-\tau(v)}
\end{equation*}
with $J_\nu(T)$ the Rayleigh-Jeans equivalent temperature, $T_\mathrm{CMB}$ the temperature of the cosmic microwave background ($T_\mathrm{CMB} = 2.73$\,K) and $T_\mathrm{c}$ the continuum temperature ($T_\mathrm{c} = 600$\,mK for NHD and $T_\mathrm{c} = 210$\,mK for ND$_2$, see Section\,\ref{continuum}).

The observed spectra were then fitted by adjusting the value of the column density, having assumed an excitation temperature. We assumed an excitation temperature of 3\,K, but the influence on the column density value is basically unchanged for an excitation temperature range of $2.8-4$\,K. This is a reasonable assumption for lines seen in absorption, however from our spectra we cannot fully exclude higher values of $T_\mathrm{ex}$, which would lead to the need for higher column densities to fit the spectra. For excitation temperatures larger than 5.3\,K for ND$_2$ and 5.8\,K for NHD, the lines would be seen in emission. For NHD, we assumed that the line is seen in absorption against a background varying with velocity and made of the dust continuum and a line possibly originating from the hot core with a FWHM linewidth of 8.5\,km\,s$^{-1}$.

With $T_\mathrm{ex} = 3$\,K, we find column densities of $N \sim  3.9\times 10^{13}$\,cm$^{-2}$ for NHD and  $N \sim 7\times 10^{12}$\,cm$^{-2}$ for para-ND$_2$. For these column density values, the maximum optical depth is 0.5 for the components of NHD and 0.3 for those of ND$_2$. Assuming 
 an ortho-to-para ratio of 0.4 for ND$_2$, consistent with the ortho-to-para ratio derived from the observations of \citet{2020arXiv200707504M},  we find that the total (ortho+para) ND$_2$ column density is $\sim  10^{13}$ cm$^{-2}$. Note that in the case of NH$_2$ the excitation temperature could be determined from the relative ratios of the hyperfine components because of the high signal-to-noise ratio of the spectra, and was found to be $\sim$\,9\,K \citep{2010A&A...521L..52H}.

We note that the critical densities associated with the detected transitions are $\gtrsim 10^6$~cm$^{-3}$ assuming collisional rate coefficients for NHD and ND$_2$ of a few 10$^{-11}$~cm$^3$s$^{-1}$ \citep[see][for the main isotopologue NH$_2$]{2019MNRAS.tmp.2223B}. Excitation temperatures are therefore expected to be lower than the kinetic temperature in the protostellar envelope where $T_\mathrm{kin}\sim 10$~K and $n \sim 10^5$~cm$^{-3}$, but will depend on the exact physical conditions of the absorbing molecules.

\begin{figure*}
  \centering
            \includegraphics[width=18cm]{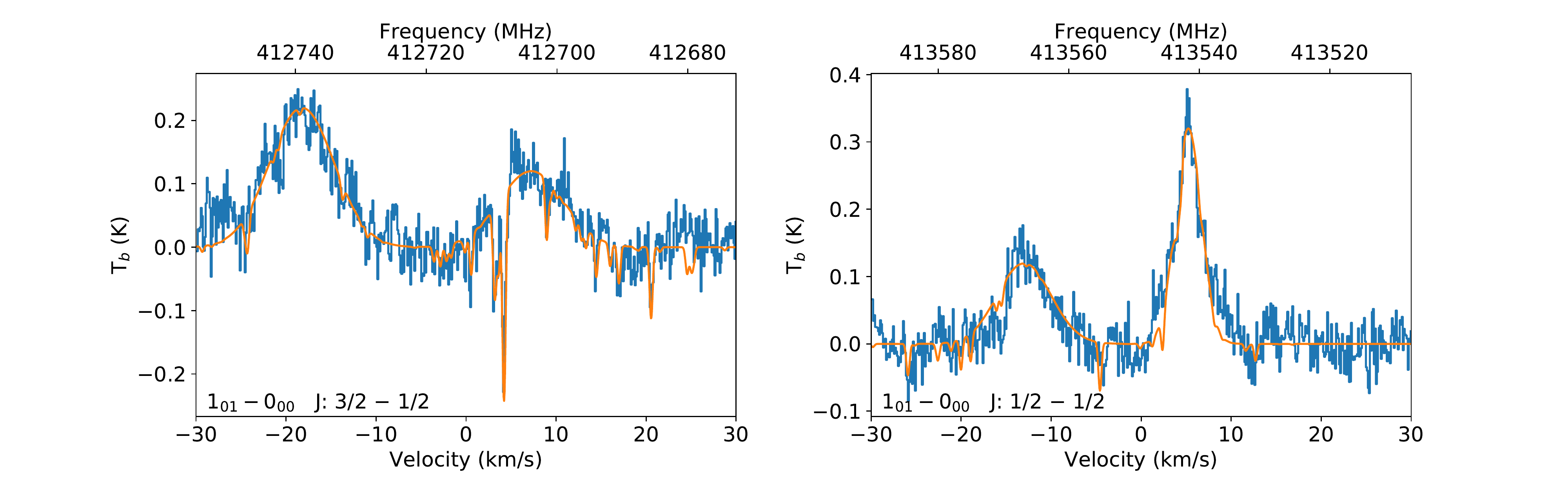}
      \caption{NHD spectra for the J: 3/2-1/2 (left panel) and J: 1/2-1/2 (right panel) transitions (blue line) and model spectra (red line). The broad emission features at 412702.5\,MHz, 412738.7\,MHz, 413541.1\,MHz, and 413564.9\,MHz, are unidentified molecular lines probably originating in the warm inner regions.}
         \label{nhdfit}
   \end{figure*}

\begin{figure}
  \centering
            \includegraphics[width=9cm]{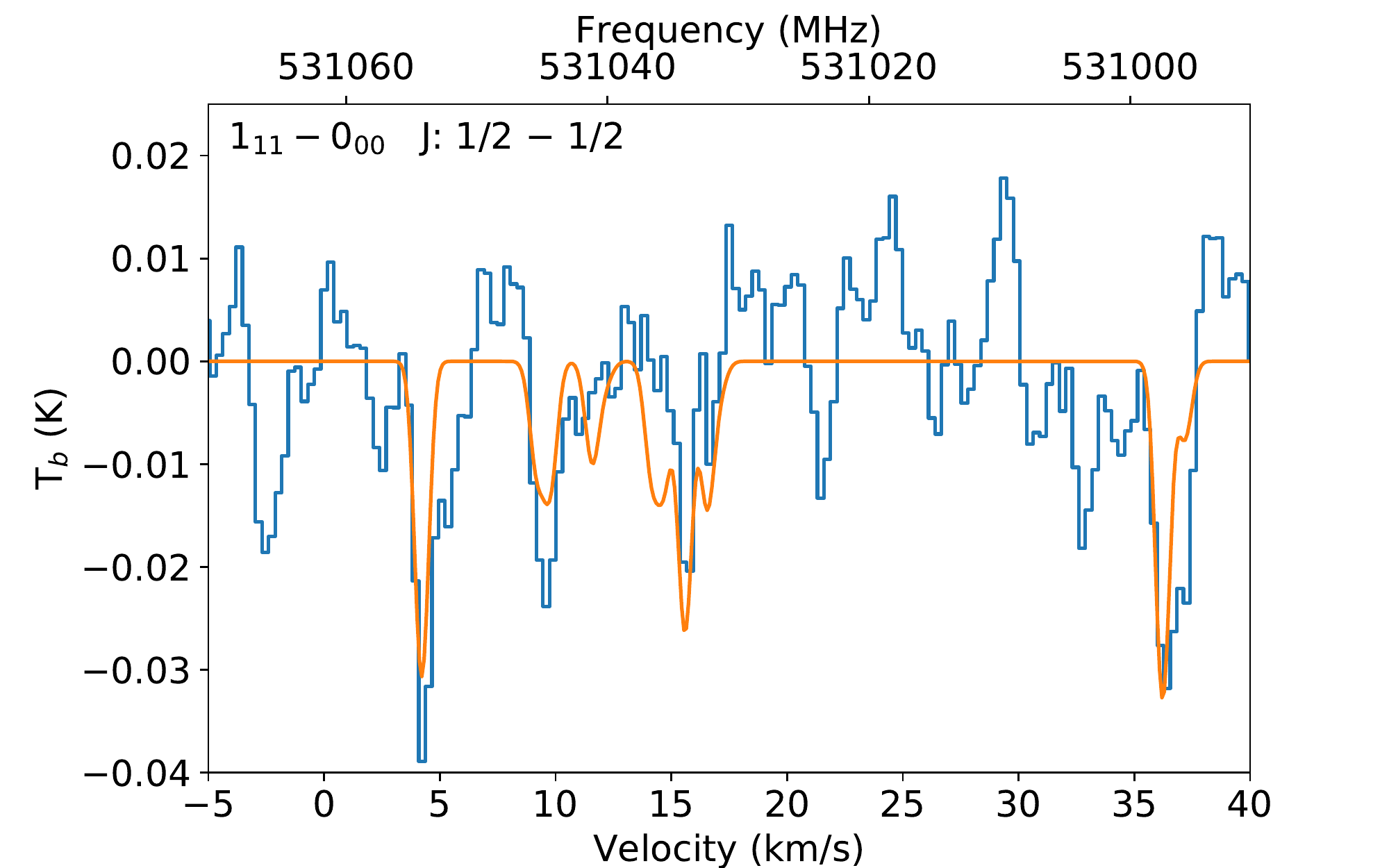}
      \caption{ND$_2$ spectrum for the J: 1/2-1/2 transition (blue line) and model spectrum (red line). }
         \label{nd2fit}
   \end{figure}

Acceptable fits are found for $N\mathrm{(NHD)} \sim 3.2 \times 10^{13} - 4.7\times 10^{13}$\,cm$^{-2}$ and we adopt this range for the uncertainty on the NHD column density value. Uncertainties on the para-ND$_2$ column density value are $\sim 5\times 10^{12} - 8\times 10^{12}$\,cm$^{-2}$. The fits are shown in Fig.\,\ref{nhdfit} and Fig.\,\ref{nd2fit} for NHD and ND$_2$, respectively.

As mentioned before, we cannot exclude that the signal against which the NHD transitions are seen is due to atmospheric fluctuations (i.e. foreground emission). Assuming this is the case, we first withdrew a high-order (e.g. 9) polynomial to the spectrum, excluding only the NHD absorption features from the fit, after which we proceeded with fitting the spectrum as described above. In this case, and assuming an excitation temperature of $T_{\rm ex}=3\,$K as well, the NHD column density needed to fit the absorption feature is $N\sim 4.6\times 10^{13}$\,cm$^{-2}$. This is within the uncertainties of our previous NHD column density determination. 

Table\,\ref{coldens} sums up the observed column densities  for ortho-NH$_2$, NHD and para-ND$_2$. These values are consistent with the ones derived by \citet{2020arXiv200707504M}. Note that for the designation of spin symmetries, we have followed the standard Maue's rule \citep{1937AnP...422..555M} where the ortho species have the largest statistical weight and para species have the smallest statistical weight. Thus, in the case of ND$_2$, the symmetric rotational levels (K$_a$+K$_c$ even) combine with antisymmetric nuclear spin function I$_D$=1 i.e. para states, while antisymmetric rotational levels (K$_a$+K$_c$ odd) combine with symmetric nuclear spin function I$_D$=0, 2 i.e. ortho states. The reverse convention was employed in \citet[][see their Appendix A.2]{2020arXiv200707504M}, hence our value of column density for para-ND$_2$ should be compared with their value for ortho-ND$_2$.

\subsection{Abundance ratios}

\citet{2010A&A...521L..52H} observed ortho-NH$_2$ at the same position in IRAS16293-2422. Based on this observation, and assuming an ortho-to-para ratio of 2, close to the value suggested by their model, \citet{2014A&A...562A..83L}  derived a total (ortho+para) NH$_2$ column density of $7.5\times 10^{13}$ cm$^{-2}$, which we adopt here. This leads to relative column density ratios NH$_2$:NHD:ND$_2$ of $\sim$  8:4:1. 

Large ratios between singly deuterated molecules and main isotopologues of a few 10\% such as we find here for NHD/NH$_2$ are common in protostellar envelopes, for molecules formed in the later stages of prestellar evolution \citep[e.g. formaldehyde, methanol, or ammonia, but unlike water, see ][]{2013A&A...550A.127T}, although the case of the amidogen radical presented here is rather extreme. For example in IRAS16293-2422, HDCO/H$_2$CO is $13-16$\% \citep{2000A&A...359.1169L}, CH$_2$DOH/CH$_3$OH is $10-50$\% \citep{2004A&A...416..159P} and NH$_2$D/NH$_3$ is 10\% \citep{1995ApJ...447..760V}. For ammonia, \citet{2005A&A...438..585R} have compiled NH$_2$D/NH$_3$ ratios between 10\% and 30\% in a sample of sources made of dense cores and protostars. Such fractionation ratios can usually be accounted for by present chemical models \citep[e.g.][]{2012ApJ...760...40A,2015A&A...576A..99R}.

The high abundance of ND$_2$  is reminiscent of other molecules, for which
 it has been observed that doubly deuterated species were often overabundant. Indeed, the ratio between the doubly- and the singly-deuterated isotopomers is in some  species even larger than the ratio between the singly deuterated  and the main isotopologue. In IRAS16293-2422, D$_2$CO/HDCO is 30$-$40\% \citep{2000A&A...359.1169L}, i.e. $2-3$ times higher than HDCO/H$_2$CO. Other protostellar envelopes are also characterised by D$_2$CO/HDCO ratios which are higher than HDCO/H$_2$CO ratios \citep{2006A&A...453..949P} \citep[see also the D$_2$CO/HDCO measurement in the cloud $\rho$ Oph A by][]{2011A&A...527A..39B}. For ammonia and methanol, the ratios XD$_2$/XHD (where X represents the species) are of the same order (though slightly lower) as the XHD/XH$_2$ ratios \citep{2006A&A...453..949P,2005A&A...438..585R}.  Water is also characterised by high abundances of its doubly deuterated isotopologue, with D$_2$O/HDO > HDO/H$_2$O in the envelope of IRAS16293-2422, as reported by \citet{2013A&A...553A..75C}. Moreover, high angular resolution interferometric observations sampling the warm inner regions of young protostars show very high abundances of doubly deuterated molecules: in NGC1333-IRAS2, D$_2$O/HDO is found to be seven times higher than HDO/H$_2$O \citep{2014ApJ...792L...5C}. Other species like formaldehyde, observed with ALMA in the hot core of IRAS16293-2422 \citep{2018A&A...610A..54P} or doubly deuterated methyl formate, which was recently detected in the hot core of IRAS16293-2422 \citep{2019A&A...623A..69M}  similarly show XD$_2$/XHD ratios typically $2-3$ times higher than the XHD/XH$_2$ ratio. So far, chemical models have struggled to explain this feature.

\section{Chemical model}
\label{mod}

We compare the observed column densities with the predictions of the gas-grain chemical model presented in \citet{2018MNRAS.477.4454H}. The model is based on the University of Grenoble Astrochemical Network (UGAN), which includes the nuclear-spin states of H$_2$, H$_2^+$, H$_3^+$ and of all the carbon, nitrogen, oxygen and sulfur containing hydrides,  
as well as their abundant deuterated forms. Grain-surface reactions are not explicitly included in the UGAN network, except the formation (and immediate desorption) of H$_2$ and isotopologues. The rates of adsorption and desorption and the list of species assumed to form in grain mantes can be found in \citet{2018MNRAS.477.4454H}. In this model, nitrogen hydrides are formed exclusively by gas-phase reactions.
An update of the oxygen hydrides chemistry was recently performed by \citet{2019MNRAS.487.3392F}. The dynamical model is presented in \citet{2018MNRAS.477.4454H} and is inspired from the studies of gravitational collapse by \citet{1969MNRAS.145..271L} and \citet{1969MNRAS.144..425P}. Briefly, 
the core, which is considered to have a constant density central plateau surrounded by an enveloppe of density falling as $r^{-2}$, where $r$ is the radius, collapses self similarly. 
 
The density of the central plateau is homogeneous and increases with time while its contribution to the total mass and total radius (and therefore total column density) decreases with time. Full details can be found in \citet{2018MNRAS.477.4454H}. The model was run for  a kinetic temperature fixed at $T=10$~K and an initial core density $n_{\rm H}=10^4$~cm$^{-3}$. Other parameters can be found in Table~1 of \citet[][`reference` model]{2018MNRAS.477.4454H}.

\begin{figure}
\centering
\includegraphics[width=8cm,angle=0]{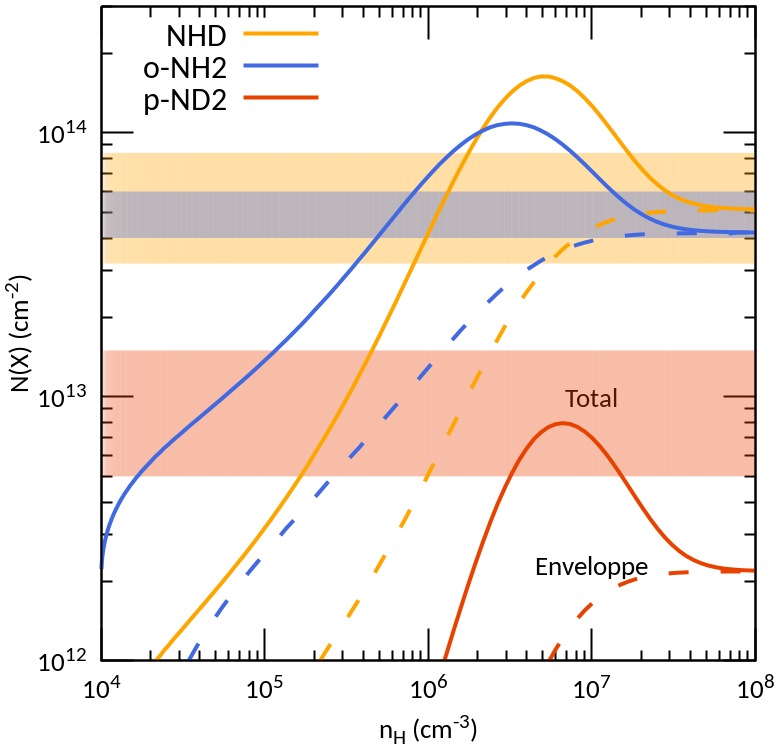}
\caption{\label{modelfigure}Model predictions \citep{2018MNRAS.477.4454H} showing the predicted column densities of NH$_2$, NHD, and ND$_2$ as a function of core density (see text). The dashed lines represent the column density of the envelope, while the solid lines represent the total (core + envelope) column density. The shaded rectangles show the observed column densities.}
         \label{model}
  \end{figure}

The total column density for ortho-NH$_2$, NHD and para-ND$_2$, and the contribution of the envelope, are shown as a function of the core density in Fig.\,\ref{modelfigure}, where the observational error bars include the uncertainty on the excitation temperatures. As may be seen, the contribution of the central plateau to the total column density reaches a maximum at a central density $n_{\rm H}\approx 3-6\times 10^6$\,cm$^{-3}$, and is then negligible above typically $5\times 10^7$\,cm$^{-3}$. Incidentally, at an evolution time of the cloud corresponding to a central density of $n_{\rm H}\approx 6.5\times10^{6}$~cm$^{-3}$, the column densities of the envelope derived from the model are $2.7\times 10^{13}$ and $2.5\times 10^{13}$\,cm$^{-2}$, for ortho-NH$_2$ and NHD, respectively. Although the model agrees with the observational constraints at this stage of cloud evolution, very good agreement is obtained also at central densities above 10$^7$\,cm$^{-3}$, while the contribution of the plateau is already mitigated. For para-ND$_2$, on the other hand, the total (core + envelope) column density matches the observation values at $n_\mathrm{H}=6.5\times 10^6$\,cm$^{-3}$. At larger core densities (where the envelope dominates) the model underestimates the observational value by a factor of $\sim 2-3$. It may be emphasised that, on observational grounds, the detected molecules most likely belong to the envelope for at least two reasons: lines are seen in absorption, and also, because the diameter of the inner plateau is $10^3$\,au, or 7\arcsec\ \citep[at a distance of 140\,pc $-$][]{2018A&A...614A..20D} when $n_{\rm H}$ reaches $6.5\times 10^5$\,cm$^{-3}$. As a result, while our model can reproduce the column density of ortho-NH$_2$ and NHD within a factor of 2, it is found to underestimate that of para-ND$_2$ by a factor larger than 2. At a central density of 10$^7$\,cm$^{-3}$, the abundance ratios are 8.7:6.4:1 for NH$_2$:NHD:ND$_2$, with ortho-NH$_2$/para-NH$_2$=1.9 and ortho-ND$_2$/para-ND$_2$=3.1. We also note that the ortho-to-para ratio of NH$_2$ stays very close to 2 throughout the collapse, while that of ND$_2$ increases steadily from 2.1 to 3.1. While the above successive deuteration ratios taking into account the total (ortho+para) column densities of NH$_2$ and ND$_2$ agree with our observations to within a factor of two (see Table\,\ref{coldens}), our model has trouble reproducing such a low ortho-to-para ratio of ND$_2$ as found by \citet{2020arXiv200707504M}. This possibly suggests that some thermalisation reactions are missing in our UGAN network, since the measured ortho-to-para ratio corresponds to a spin temperature as low as $5-6$\,K\footnote{We note that at temperatures below $\sim$ 10\,K, the ortho-to-para ratio of ND$_2$ can be approximated by the function $6\times\exp{(-15.5/T_\mathrm{spin})}$}. Finally, our model predicts that the NH$_2$:NHD:ND$_2$ ratios remain constant and equal to 7.1:5.6:1 above a density of $3\times 10^7$ cm$^{-3}$.
	
These results indicate that pure gas-phase chemistry as in the UGAN model can reproduce the observed column densities and abundance ratios of the amidogen radical isotopologues in protostellar envelopes within a factor of $\sim$\,2$-$3. We note, however, that we have assumed identical rate coefficients for dissociative recombination of all deuterated isotopologues of NH$_4^+$ with, in addition, statistical H/D branching ratios for the products. These assumptions are questionable because DR experiments have suggested the occurence of isotope effects \citep[e.g.][]{2004JChPh.120.7391O}. This will be investigated in a future, more complete, modelling study.

\begin{table}
	\center
	\caption{Comparison  between chemical model predictions at a central density of 10$^7$ cm$^{-3}$, and observational constraints for the column density of ortho-NH$_2$, NHD, and para-ND$_2$. The model values are for the envelope. Also given are the ortho-to-para ratios and total column densities for NH$_2$ and ND$_2$ as obtained with the UGAN model. The last column sums up the observational values from the work by \citet{2020arXiv200707504M}. \label{coldens}}
	\begin{tabular}{c c c c}
		Species & Model & Observation & \citet{2020arXiv200707504M} \\
		& & (this work) & \\
		& (cm$^{-2}$) & (cm$^{-2}$) & (cm$^{-2}$) \\
		\hline
		ortho-NH$_2$ &            $2.9\times 10^{13}$& $5.0\times 10^{13}$ & $5.4\times 10^{13}$\\
		para-NH$_2$ &            $1.5\times 10^{13}$\\
		NHD  &            $3.2\times 10^{13}$ & $3.9\times 10^{13}$ & $4.7\times 10^{13}$ \\
		ortho-ND$_2$ &            $3.8\times 10^{12}$ & & $2.4\times 10^{12}$ \\
		para-ND$_2$ &            $1.2\times 10^{12}$ & $7.0\times 10^{12}$ & $6.6\times 10^{12}$ \\
		NH$_2$  & $4.4\times 10^{13}$\\
		ND$_2$  & $5.0\times 10^{12}$ \\
		ortho-NH$_2$/para-NH$_2$ &     1.9\\
		ortho-ND$_2$/para-ND$_2$ &     3.1 & & 0.4\\
		NH$_2$/ND$_2$   &      8.7\\
		NHD/ND$_2$   &      6.4\\
		\hline
	\end{tabular}
\end{table}

Statistical distributions of deuterium atoms is expected if nitrogen hydrides form by successive hydrogenations of nitrogen atoms on grain surfaces \citep{1982A&A...114..245T,1989MNRAS.240P..25B,1997ApJ...482L.203C}. 
According to the grain-surface addition scheme, NHD/NH$_2 = 2$\,D/H (where D/H denotes the {\it atomic} D over H ratio) and ND$_2$/NH$_2$= (D/H)$^2$. This leads to ND$_2$/NHD =  0.25 NHD/NH$_2$. This value is  consistent  with the observed ratios, for which ND$_2$/NHD = 1/2 NHD/NH$_2$ within a factor of two. 
Therefore, although grain surface chemistry is not needed to account for the observed abundances of these species, we cannot exclude grain-surface processes either. 
We note that the ortho-to-para ratios provide additional constraints but current measurements have large uncertainties \citep{2020arXiv200707504M,2017A&A...600A..61H,2018MNRAS.477.4454H}. Indeed, taking the ortho-to-para ratio for ND$_2$ as 3 as indicated by our chemical model in the late collapse stages, we find a total ND$_2$ column density of $4.8\times 10^{13}$\,cm$^{-2}$. This in turn leads to ND$_2$/NHD = 4/3 NHD/NH$_2$, with which the statistical grain-surface scenario is inconsistent.

Our analysis and chemical model are based on the hypothesis that all species are coexistent. However, in the case of NH and ND, a detailed radiative transfer model of the spectra of these two species seen towards the dense core IRAS16293E showed that the signal from NH and that from ND came from spatially separate regions \citep{2016A&A...587A..26B}, with the ND transition sampling higher density material. 
It is possible as well that in the case of amidogen, the deuterated isotopologue is concentrated at higher densities than NH$_2$, and even more so for the doubly-deuterated isotopologue. Indeed deuteration proceeds faster at higher densities because it is tightly linked to the depletion of abundant gas-phase species like CO. 
The discrepancy between the model and the ND$_2$ observations could also indicate that important fractionation reactions are missing from the model.

In order to assess the sensitivity of our results to the assumed  excitation temperature, we have derived the column densities assuming $T_{\rm ex}=4.5$\,K. With this value, we find $N$(NHD)$ = 7\times 10^{13}$\,cm$^{-2}$ and $N$(para-ND$_2) = 1.2\times 10^{13}$\,cm$^{-2}$, i.e. a total (ortho+para) column density of ND$_2$ $N$(ND$_2$) =  $1.7\times 10^{13}$\,cm$^{-2}$, assuming as before an ortho-to-para ratio of 0.4 for ND$_2$. While for NH$_2$, the excitation temperature could be determined by fitting the relative ratios of the hyperfine components \citep{2010A&A...521L..52H}, this is not possible for NHD and ND$_2$ because of the low signal-to-noise ratio of the observations. With these column density values, the relative ratios between the different amidogen isotopologues are  4 :  4 : 1 for NH$_2$ : NHD : ND$_2$. Clearly, a non-LTE model is required in order to derive more accurate column densities and abundance ratios. This will be performed in a forthcoming paper thanks to the next availability of collisional rate coefficients for NHD and ND$_2$ (Bop et al., in prep.) 

\section{Conclusions}
\label{conclu}
We have detected the singly-deuterated  and doubly-deuterated isotopologues of amidogen, NHD and para-ND$_2$  towards the envelope of the class 0 protostar IRAS16293-2422. The column densities of NHD and  para-ND$_2$ are estimated to be $3.2 \times 10^{13} - 7\times 10^{13}$ cm$^{-2}$ and  $5\times 10^{12} - 1.2\times 10^{13}$ cm$^{-2}$, respectively. The
observed relative ratios between the deuterated isotopologues and the main isotopologue NH$_2$ can be reproduced within a factor of  $2-3$ with our pure gas-phase chemical model. 
This result adds to previous evidence that pure gas-phase chemistry can reproduce many features
(i.e. spin-state ratios for NH$_2$, deuterium fractionation, abundances) observed in nitrogen hydrides \citep{2018MNRAS.477.4454H}. However, the data are also
consistent  with the statistical ratios expected from grain-surface chemistry. A more elaborate analysis of the observations is necessary taking into account the different angular resolutions for NHD and ND$_2$ as well as likely non-LTE excitation effects. A better understanding  of the ortho-to-para ratio in ND$_2$ might help to distinguish between both chemical routes.

\section*{Acknowledgements}
This research has made 
use of data from the {\it Herschel} Gould Belt survey (HGBS) project (http://gouldbelt-herschel.cea.fr). The HGBS is a {\it Herschel} Key Programme jointly 
carried out by SPIRE Specialist Astronomy Group 3 (SAG 3), scientists of several institutes in the PACS Consortium 
(CEA Saclay, INAF-IFSI Rome and INAF-Arcetri, KU Leuven, MPIA Heidelberg), and scientists of the {\it Herschel} Science Center (HSC). This work has been supported by the Agence Nationale de la Recherche (ANR-HYDRIDES), contract ANR-12-BS05-0011-01. This work was supported by the Programme National "Physique et
Chimie du Milieu Interstellaire" (PCMI) of CNRS/INSU with INC/INP co-funded
by CEA and CNES.

\section*{Data Availability}

The data underlying this article were accessed from the ESO archive (APEX data) and The Herschel Space Observatory archive (HIFI/Herschel data).



\bibliographystyle{mnras}
\bibliography{biblio} 

\begin{thebibliography}{}
\makeatletter
\relax
\def\mn@urlcharsother{\let\do\@makeother \do\$\do\&\do\#\do\^\do\_\do\%\do\~}
\def\mn@doi{\begingroup\mn@urlcharsother \@ifnextchar [ {\mn@doi@}
  {\mn@doi@[]}}
\def\mn@doi@[#1]#2{\def\@tempa{#1}\ifx\@tempa\@empty \href
  {http://dx.doi.org/#2} {doi:#2}\else \href {http://dx.doi.org/#2} {#1}\fi
  \endgroup}
\def\mn@eprint#1#2{\mn@eprint@#1:#2::\@nil}
\def\mn@eprint@arXiv#1{\href {http://arxiv.org/abs/#1} {{\tt arXiv:#1}}}
\def\mn@eprint@dblp#1{\href {http://dblp.uni-trier.de/rec/bibtex/#1.xml}
  {dblp:#1}}
\def\mn@eprint@#1:#2:#3:#4\@nil{\def\@tempa {#1}\def\@tempb {#2}\def\@tempc
  {#3}\ifx \@tempc \@empty \let \@tempc \@tempb \let \@tempb \@tempa \fi \ifx
  \@tempb \@empty \def\@tempb {arXiv}\fi \@ifundefined
  {mn@eprint@\@tempb}{\@tempb:\@tempc}{\expandafter \expandafter \csname
  mn@eprint@\@tempb\endcsname \expandafter{\@tempc}}}

\bibitem[\protect\citeauthoryear{{Aikawa}, {Wakelam}, {Hersant}, {Garrod}  \&
  {Herbst}}{{Aikawa} et~al.}{2012}]{2012ApJ...760...40A}
{Aikawa} Y.,  {Wakelam} V.,  {Hersant} F.,  {Garrod} R.~T.,   {Herbst} E.,
  2012, \mn@doi [\apj] {10.1088/0004-637X/760/1/40}, \href
  {http://adsabs.harvard.edu/abs/2012ApJ...760...40A} {760, 40}

\bibitem[\protect\citeauthoryear{{Andr{\'e}} et~al.,}{{Andr{\'e}}
  et~al.}{2010}]{2010A&A...518L.102A}
{Andr{\'e}} P.,  et~al., 2010, \mn@doi [\aap] {10.1051/0004-6361/201014666},
  \href {http://cdsads.u-strasbg.fr/abs/2010A%26A...518L.102A} {518, L102}

\bibitem[\protect\citeauthoryear{{Aniano}, {Draine}, {Gordon}  \&
  {Sandstrom}}{{Aniano} et~al.}{2011}]{2011PASP..123.1218A}
{Aniano} G.,  {Draine} B.~T.,  {Gordon} K.~D.,   {Sandstrom} K.,  2011, \mn@doi
  [\pasp] {10.1086/662219}, \href
  {http://adsabs.harvard.edu/abs/2011PASP..123.1218A} {123, 1218}

\bibitem[\protect\citeauthoryear{{Bacmann} et~al.,}{{Bacmann}
  et~al.}{2010}]{2010A&A...521L..42B}
{Bacmann} A.,  et~al., 2010, \mn@doi [\aap] {10.1051/0004-6361/201015102},
  \href {http://adsabs.harvard.edu/abs/2010A%26A...521L..42B} {521, L42}

\bibitem[\protect\citeauthoryear{{Bacmann} et~al.,}{{Bacmann}
  et~al.}{2016}]{2016A&A...587A..26B}
{Bacmann} A.,  et~al., 2016, \mn@doi [\aap] {10.1051/0004-6361/201526084},
  \href {http://adsabs.harvard.edu/abs/2016A%26A...587A..26B} {587, A26}

\bibitem[\protect\citeauthoryear{{Bergman}, {Parise}, {Liseau}  \&
  {Larsson}}{{Bergman} et~al.}{2011}]{2011A&A...527A..39B}
{Bergman} P.,  {Parise} B.,  {Liseau} R.,   {Larsson} B.,  2011, \mn@doi [\aap]
  {10.1051/0004-6361/201015012}, \href
  {http://adsabs.harvard.edu/abs/2011A%26A...527A..39B} {527, A39}

\bibitem[\protect\citeauthoryear{{Bouhafs}, {Bacmann}, {Faure}  \&
  {Lique}}{{Bouhafs} et~al.}{2019}]{2019MNRAS.tmp.2223B}
{Bouhafs} N.,  {Bacmann} A.,  {Faure} A.,   {Lique} F.,  2019, \mn@doi [\mnras]
  {10.1093/mnras/stz2586}, \href
  {https://ui.adsabs.harvard.edu/abs/2019MNRAS.tmp.2223B} {p.~2223}

\bibitem[\protect\citeauthoryear{{Brown} \& {Millar}}{{Brown} \&
  {Millar}}{1989}]{1989MNRAS.240P..25B}
{Brown} P.~D.,  {Millar} T.~J.,  1989, \mn@doi [\mnras]
  {10.1093/mnras/240.1.25P}, \href
  {http://adsabs.harvard.edu/abs/1989MNRAS.240P..25B} {240, 25P}

\bibitem[\protect\citeauthoryear{{Castets}, {Ceccarelli}, {Loinard}, {Caux}  \&
  {Lefloch}}{{Castets} et~al.}{2001}]{2001A&A...375...40C}
{Castets} A.,  {Ceccarelli} C.,  {Loinard} L.,  {Caux} E.,   {Lefloch} B.,
  2001, \mn@doi [\aap] {10.1051/0004-6361:20010662}, \href
  {http://adsabs.harvard.edu/abs/2001A%26A...375...40C} {375, 40}

\bibitem[\protect\citeauthoryear{{Caux} et~al.,}{{Caux}
  et~al.}{2011}]{2011A&A...532A..23C}
{Caux} E.,  et~al., 2011, \mn@doi [\aap] {10.1051/0004-6361/201015399}, \href
  {https://ui.adsabs.harvard.edu/abs/2011A&A...532A..23C} {532, A23}

\bibitem[\protect\citeauthoryear{{Charnley}, {Tielens}  \&
  {Rodgers}}{{Charnley} et~al.}{1997}]{1997ApJ...482L.203C}
{Charnley} S.~B.,  {Tielens} A.~G.~G.~M.,   {Rodgers} S.~D.,  1997, \mn@doi
  [\apjl] {10.1086/310697}, \href
  {http://adsabs.harvard.edu/abs/1997ApJ...482L.203C} {482, L203}

\bibitem[\protect\citeauthoryear{{Coutens} et~al.,}{{Coutens}
  et~al.}{2013}]{2013A&A...553A..75C}
{Coutens} A.,  et~al., 2013, \mn@doi [\aap] {10.1051/0004-6361/201220967},
  \href {http://adsabs.harvard.edu/abs/2013A%26A...553A..75C} {553, A75}

\bibitem[\protect\citeauthoryear{{Coutens}, {J{\o}rgensen}, {Persson}, {van
  Dishoeck}, {Vastel}  \& {Taquet}}{{Coutens}
  et~al.}{2014}]{2014ApJ...792L...5C}
{Coutens} A.,  {J{\o}rgensen} J.~K.,  {Persson} M.~V.,  {van Dishoeck} E.~F.,
  {Vastel} C.,   {Taquet} V.,  2014, \mn@doi [\apjl]
  {10.1088/2041-8205/792/1/L5}, \href
  {http://adsabs.harvard.edu/abs/2014ApJ...792L...5C} {792, L5}

\bibitem[\protect\citeauthoryear{{Dumouchel}, {K{\l}os}, {Tobo{\l}a},
  {Bacmann}, {Maret}, {Hily-Blant}, {Faure}  \& {Lique}}{{Dumouchel}
  et~al.}{2012}]{2012JChPh.137k4306D}
{Dumouchel} F.,  {K{\l}os} J.,  {Tobo{\l}a} R.,  {Bacmann} A.,  {Maret} S.,
  {Hily-Blant} P.,  {Faure} A.,   {Lique} F.,  2012, \mn@doi [\jcp]
  {10.1063/1.4753423}, \href
  {https://ui.adsabs.harvard.edu/abs/2012JChPh.137k4306D} {137, 114306}

\bibitem[\protect\citeauthoryear{{Dzib} et~al.,}{{Dzib}
  et~al.}{2018}]{2018A&A...614A..20D}
{Dzib} S.~A.,  et~al., 2018, \mn@doi [\aap] {10.1051/0004-6361/201732093},
  \href {https://ui.adsabs.harvard.edu/abs/2018A&A...614A..20D} {614, A20}

\bibitem[\protect\citeauthoryear{{Faure}, {Hily-Blant}, {Rist}, {Pineau des
  For{\^e}ts}, {Matthews}  \& {Flower}}{{Faure}
  et~al.}{2019}]{2019MNRAS.487.3392F}
{Faure} A.,  {Hily-Blant} P.,  {Rist} C.,  {Pineau des For{\^e}ts} G.,
  {Matthews} A.,   {Flower} D.~R.,  2019, \mn@doi [\mnras]
  {10.1093/mnras/stz1531}, \href
  {https://ui.adsabs.harvard.edu/abs/2019MNRAS.487.3392F} {487, 3392}

\bibitem[\protect\citeauthoryear{{Griffin} et~al.,}{{Griffin}
  et~al.}{2010}]{2010A&A...518L...3G}
{Griffin} M.~J.,  et~al., 2010, \mn@doi [\aap] {10.1051/0004-6361/201014519},
  \href {http://adsabs.harvard.edu/abs/2010A%26A...518L...3G} {518, L3}

\bibitem[\protect\citeauthoryear{{G{\"u}sten}, {Nyman}, {Schilke}, {Menten},
  {Cesarsky}  \& {Booth}}{{G{\"u}sten} et~al.}{2006}]{2006A&A...454L..13G}
{G{\"u}sten} R.,  {Nyman} L.~{\AA}.,  {Schilke} P.,  {Menten} K.,  {Cesarsky}
  C.,   {Booth} R.,  2006, \mn@doi [\aap] {10.1051/0004-6361:20065420}, \href
  {http://adsabs.harvard.edu/abs/2006A%26A...454L..13G} {454, L13}

\bibitem[\protect\citeauthoryear{{Harju} et~al.,}{{Harju}
  et~al.}{2017}]{2017A&A...600A..61H}
{Harju} J.,  et~al., 2017, \mn@doi [\aap] {10.1051/0004-6361/201628463}, \href
  {https://ui.adsabs.harvard.edu/abs/2017A&A...600A..61H} {600, A61}

\bibitem[\protect\citeauthoryear{{Hily-Blant} et~al.,}{{Hily-Blant}
  et~al.}{2010}]{2010A&A...521L..52H}
{Hily-Blant} P.,  et~al., 2010, \mn@doi [\aap] {10.1051/0004-6361/201015253},
  \href {http://adsabs.harvard.edu/abs/2010A%26A...521L..52H} {521, L52}

\bibitem[\protect\citeauthoryear{{Hily-Blant}, {Faure}, {Rist}, {Pineau des
  For{\^e}ts}  \& {Flower}}{{Hily-Blant} et~al.}{2018}]{2018MNRAS.477.4454H}
{Hily-Blant} P.,  {Faure} A.,  {Rist} C.,  {Pineau des For{\^e}ts} G.,
  {Flower} D.~R.,  2018, \mn@doi [\mnras] {10.1093/mnras/sty881}, \href
  {http://adsabs.harvard.edu/abs/2018MNRAS.477.4454H} {477, 4454}

\bibitem[\protect\citeauthoryear{{Holland} et~al.,}{{Holland}
  et~al.}{2013}]{2013MNRAS.430.2513H}
{Holland} W.~S.,  et~al., 2013, \mn@doi [\mnras] {10.1093/mnras/sts612}, \href
  {http://cdsads.u-strasbg.fr/abs/2013MNRAS.430.2513H} {430, 2513}

\bibitem[\protect\citeauthoryear{{J{\o}rgensen}, {Bourke}, {Nguyen Luong}  \&
  {Takakuwa}}{{J{\o}rgensen} et~al.}{2011}]{2011A&A...534A.100J}
{J{\o}rgensen} J.~K.,  {Bourke} T.~L.,  {Nguyen Luong} Q.,   {Takakuwa} S.,
  2011, \mn@doi [\aap] {10.1051/0004-6361/201117139}, \href
  {https://ui.adsabs.harvard.edu/abs/2011A&A...534A.100J} {534, A100}

\bibitem[\protect\citeauthoryear{{J{\o}rgensen} et~al.,}{{J{\o}rgensen}
  et~al.}{2016}]{2016A&A...595A.117J}
{J{\o}rgensen} J.~K.,  et~al., 2016, \mn@doi [\aap]
  {10.1051/0004-6361/201628648}, \href
  {http://adsabs.harvard.edu/abs/2016A%26A...595A.117J} {595, A117}

\bibitem[\protect\citeauthoryear{{Kanada}, {Yamamoto}  \& {Saito}}{{Kanada}
  et~al.}{1991}]{1991JChPh..94.3423K}
{Kanada} M.,  {Yamamoto} S.,   {Saito} S.,  1991, \mn@doi [\jcp]
  {10.1063/1.460706}, \href {http://adsabs.harvard.edu/abs/1991JChPh..94.3423K}
  {94, 3423}

\bibitem[\protect\citeauthoryear{{Kobayashi}, {Ozeki}, {Saito}, {Tonooka}  \&
  {Yamamoto}}{{Kobayashi} et~al.}{1997}]{1997JChPh.107.9289K}
{Kobayashi} K.,  {Ozeki} H.,  {Saito} S.,  {Tonooka} M.,   {Yamamoto} S.,
  1997, \mn@doi [\jcp] {10.1063/1.475224}, \href
  {http://adsabs.harvard.edu/abs/1997JChPh.107.9289K} {107, 9289}

\bibitem[\protect\citeauthoryear{{Larson}}{{Larson}}{1969}]{1969MNRAS.145..271L}
{Larson} R.~B.,  1969, \mn@doi [\mnras] {10.1093/mnras/145.3.271}, \href
  {https://ui.adsabs.harvard.edu/abs/1969MNRAS.145..271L} {145, 271}

\bibitem[\protect\citeauthoryear{{Le Gal}, {Hily-Blant}, {Faure}, {Pineau des
  For{\^e}ts}, {Rist}  \& {Maret}}{{Le Gal} et~al.}{2014}]{2014A&A...562A..83L}
{Le Gal} R.,  {Hily-Blant} P.,  {Faure} A.,  {Pineau des For{\^e}ts} G.,
  {Rist} C.,   {Maret} S.,  2014, \mn@doi [\aap] {10.1051/0004-6361/201322386},
  \href {http://adsabs.harvard.edu/abs/2014A%26A...562A..83L} {562, A83}

\bibitem[\protect\citeauthoryear{{Linsky} et~al.,}{{Linsky}
  et~al.}{2006}]{2006ApJ...647.1106L}
{Linsky} J.~L.,  et~al., 2006, \mn@doi [\apj] {10.1086/505556}, \href
  {http://adsabs.harvard.edu/abs/2006ApJ...647.1106L} {647, 1106}

\bibitem[\protect\citeauthoryear{{Lis}, {Gerin}, {Phillips}  \& {Motte}}{{Lis}
  et~al.}{2002a}]{2002ApJ...569..322L}
{Lis} D.~C.,  {Gerin} M.,  {Phillips} T.~G.,   {Motte} F.,  2002a, \mn@doi
  [\apj] {10.1086/339232}, \href
  {http://adsabs.harvard.edu/abs/2002ApJ...569..322L} {569, 322}

\bibitem[\protect\citeauthoryear{{Lis}, {Roueff}, {Gerin}, {Phillips},
  {Coudert}, {van der Tak}  \& {Schilke}}{{Lis}
  et~al.}{2002b}]{2002ApJ...571L..55L}
{Lis} D.~C.,  {Roueff} E.,  {Gerin} M.,  {Phillips} T.~G.,  {Coudert} L.~H.,
  {van der Tak} F.~F.~S.,   {Schilke} P.,  2002b, \mn@doi [\apjl]
  {10.1086/341132}, \href {http://adsabs.harvard.edu/abs/2002ApJ...571L..55L}
  {571, L55}

\bibitem[\protect\citeauthoryear{{Loinard}, {Castets}, {Ceccarelli}, {Tielens},
  {Faure}, {Caux}  \& {Duvert}}{{Loinard} et~al.}{2000}]{2000A&A...359.1169L}
{Loinard} L.,  {Castets} A.,  {Ceccarelli} C.,  {Tielens} A.~G.~G.~M.,  {Faure}
  A.,  {Caux} E.,   {Duvert} G.,  2000, \aap, \href
  {http://adsabs.harvard.edu/abs/2000A%26A...359.1169L} {359, 1169}

\bibitem[\protect\citeauthoryear{{Manigand} et~al.,}{{Manigand}
  et~al.}{2019}]{2019A&A...623A..69M}
{Manigand} S.,  et~al., 2019, \mn@doi [\aap] {10.1051/0004-6361/201832844},
  \href {https://ui.adsabs.harvard.edu/abs/2019A&A...623A..69M} {623, A69}

\bibitem[\protect\citeauthoryear{{Maue}}{{Maue}}{1937}]{1937AnP...422..555M}
{Maue} A.~W.,  1937, \mn@doi [Annalen der Physik] {10.1002/andp.19374220608},
  \href {https://ui.adsabs.harvard.edu/abs/1937AnP...422..555M} {422, 555}

\bibitem[\protect\citeauthoryear{{Melosso}, {Degli Esposti}  \&
  {Dore}}{{Melosso} et~al.}{2017}]{2017ApJS..233...15M}
{Melosso} M.,  {Degli Esposti} C.,   {Dore} L.,  2017, \mn@doi [\apjs]
  {10.3847/1538-4365/aa9220}, \href
  {https://ui.adsabs.harvard.edu/abs/2017ApJS..233...15M} {233, 15}

\bibitem[\protect\citeauthoryear{{Melosso} et~al.,}{{Melosso}
  et~al.}{2020}]{2020arXiv200707504M}
{Melosso} M.,  et~al., 2020, arXiv e-prints, \href
  {https://ui.adsabs.harvard.edu/abs/2020arXiv200707504M} {p. arXiv:2007.07504}

\bibitem[\protect\citeauthoryear{{Motoki}, {Ozeki}  \& {Kobayashi}}{{Motoki}
  et~al.}{2013}]{2013mss..confETH01M}
{Motoki} Y.,  {Ozeki} H.,   {Kobayashi} K.,  2013, in 68th International
  Symposium on Molecular Spectroscopy. p. ETH01

\bibitem[\protect\citeauthoryear{{{\"O}jekull} et~al.,}{{{\"O}jekull}
  et~al.}{2004}]{2004JChPh.120.7391O}
{{\"O}jekull} J.,  et~al., 2004, \mn@doi [\jcp] {10.1063/1.1669388}, \href
  {https://ui.adsabs.harvard.edu/abs/2004JChPh.120.7391O} {120, 7391}

\bibitem[\protect\citeauthoryear{{Ott}}{{Ott}}{2010}]{2010ASPC..434..139O}
{Ott} S.,  2010, in {Mizumoto} Y.,  {Morita} K.~I.,   {Ohishi} M.,  eds,
  Astronomical Society of the Pacific Conference Series Vol. 434, Astronomical
  Data Analysis Software and Systems XIX. p.~139 (\mn@eprint {arXiv}
  {1011.1209})

\bibitem[\protect\citeauthoryear{{Pagani}, {Salez}  \& {Wannier}}{{Pagani}
  et~al.}{1992}]{1992A&A...258..479P}
{Pagani} L.,  {Salez} M.,   {Wannier} P.~G.,  1992, \aap, \href
  {http://adsabs.harvard.edu/abs/1992A%26A...258..479P} {258, 479}

\bibitem[\protect\citeauthoryear{{Parise}, {Castets}, {Herbst}, {Caux},
  {Ceccarelli}, {Mukhopadhyay}  \& {Tielens}}{{Parise}
  et~al.}{2004}]{2004A&A...416..159P}
{Parise} B.,  {Castets} A.,  {Herbst} E.,  {Caux} E.,  {Ceccarelli} C.,
  {Mukhopadhyay} I.,   {Tielens} A.~G.~G.~M.,  2004, \mn@doi [\aap]
  {10.1051/0004-6361:20034490}, \href
  {http://adsabs.harvard.edu/abs/2004A%26A...416..159P} {416, 159}

\bibitem[\protect\citeauthoryear{{Parise}, {Ceccarelli}, {Tielens}, {Castets},
  {Caux}, {Lefloch}  \& {Maret}}{{Parise} et~al.}{2006}]{2006A&A...453..949P}
{Parise} B.,  {Ceccarelli} C.,  {Tielens} A.~G.~G.~M.,  {Castets} A.,  {Caux}
  E.,  {Lefloch} B.,   {Maret} S.,  2006, \mn@doi [\aap]
  {10.1051/0004-6361:20054476}, \href
  {http://adsabs.harvard.edu/abs/2006A%26A...453..949P} {453, 949}

\bibitem[\protect\citeauthoryear{{Pattle} et~al.,}{{Pattle}
  et~al.}{2015}]{2015MNRAS.450.1094P}
{Pattle} K.,  et~al., 2015, \mn@doi [\mnras] {10.1093/mnras/stv376}, \href
  {http://adsabs.harvard.edu/abs/2015MNRAS.450.1094P} {450, 1094}

\bibitem[\protect\citeauthoryear{{Penston}}{{Penston}}{1969}]{1969MNRAS.144..425P}
{Penston} M.~V.,  1969, \mn@doi [\mnras] {10.1093/mnras/144.4.425}, \href
  {https://ui.adsabs.harvard.edu/abs/1969MNRAS.144..425P} {144, 425}

\bibitem[\protect\citeauthoryear{{Persson} et~al.,}{{Persson}
  et~al.}{2018}]{2018A&A...610A..54P}
{Persson} M.~V.,  et~al., 2018, \mn@doi [\aap] {10.1051/0004-6361/201731684},
  \href {http://adsabs.harvard.edu/abs/2018A%26A...610A..54P} {610, A54}

\bibitem[\protect\citeauthoryear{{Pilbratt} et~al.,}{{Pilbratt}
  et~al.}{2010}]{2010A&A...518L...1P}
{Pilbratt} G.~L.,  et~al., 2010, \mn@doi [\aap] {10.1051/0004-6361/201014759},
  \href {http://adsabs.harvard.edu/abs/2010A%26A...518L...1P} {518, L1}

\bibitem[\protect\citeauthoryear{{Roberts}, {Herbst}  \& {Millar}}{{Roberts}
  et~al.}{2003}]{2003ApJ...591L..41R}
{Roberts} H.,  {Herbst} E.,   {Millar} T.~J.,  2003, \mn@doi [\apjl]
  {10.1086/376962}, \href {http://adsabs.harvard.edu/abs/2003ApJ...591L..41R}
  {591, L41}

\bibitem[\protect\citeauthoryear{{Roueff}, {Lis}, {van der Tak}, {Gerin}  \&
  {Goldsmith}}{{Roueff} et~al.}{2005}]{2005A&A...438..585R}
{Roueff} E.,  {Lis} D.~C.,  {van der Tak} F.~F.~S.,  {Gerin} M.,   {Goldsmith}
  P.~F.,  2005, \mn@doi [\aap] {10.1051/0004-6361:20052724}, \href
  {http://adsabs.harvard.edu/abs/2005A%26A...438..585R} {438, 585}

\bibitem[\protect\citeauthoryear{{Roueff}, {Loison}  \& {Hickson}}{{Roueff}
  et~al.}{2015}]{2015A&A...576A..99R}
{Roueff} E.,  {Loison} J.~C.,   {Hickson} K.~M.,  2015, \mn@doi [\aap]
  {10.1051/0004-6361/201425113}, \href
  {http://adsabs.harvard.edu/abs/2015A%26A...576A..99R} {576, A99}

\bibitem[\protect\citeauthoryear{{Shipman} et~al.,}{{Shipman}
  et~al.}{2017}]{2017A&A...608A..49S}
{Shipman} R.~F.,  et~al., 2017, \mn@doi [\aap] {10.1051/0004-6361/201731385},
  \href {http://adsabs.harvard.edu/abs/2017A%26A...608A..49S} {608, A49}

\bibitem[\protect\citeauthoryear{{Taquet}, {Peters}, {Kahane}, {Ceccarelli},
  {L{\'o}pez-Sepulcre}, {Toubin}, {Duflot}  \& {Wiesenfeld}}{{Taquet}
  et~al.}{2013}]{2013A&A...550A.127T}
{Taquet} V.,  {Peters} P.~S.,  {Kahane} C.,  {Ceccarelli} C.,
  {L{\'o}pez-Sepulcre} A.,  {Toubin} C.,  {Duflot} D.,   {Wiesenfeld} L.,
  2013, \mn@doi [\aap] {10.1051/0004-6361/201220084}, \href
  {http://adsabs.harvard.edu/abs/2013A%26A...550A.127T} {550, A127}

\bibitem[\protect\citeauthoryear{{Tielens} \& {Hagen}}{{Tielens} \&
  {Hagen}}{1982}]{1982A&A...114..245T}
{Tielens} A.~G.~G.~M.,  {Hagen} W.,  1982, A\&A, \href
  {http://adsabs.harvard.edu/abs/1982A%26A...114..245T} {114, 245}

\bibitem[\protect\citeauthoryear{{Vassilev} et~al.,}{{Vassilev}
  et~al.}{2008}]{2008A&A...490.1157V}
{Vassilev} V.,  et~al., 2008, \mn@doi [\aap] {10.1051/0004-6361:200810459},
  \href {http://adsabs.harvard.edu/abs/2008A%26A...490.1157V} {490, 1157}

\bibitem[\protect\citeauthoryear{{Walmsley}, {Flower}  \& {Pineau des
  For{\^e}ts}}{{Walmsley} et~al.}{2004}]{2004A&A...418.1035W}
{Walmsley} C.~M.,  {Flower} D.~R.,   {Pineau des For{\^e}ts} G.,  2004, \mn@doi
  [\aap] {10.1051/0004-6361:20035718}, \href
  {http://adsabs.harvard.edu/abs/2004A%26A...418.1035W} {418, 1035}

\bibitem[\protect\citeauthoryear{{de Graauw} et~al.,}{{de Graauw}
  et~al.}{2010}]{2010A&A...518L...6D}
{de Graauw} T.,  et~al., 2010, \mn@doi [\aap] {10.1051/0004-6361/201014698},
  \href {http://adsabs.harvard.edu/abs/2010A%26A...518L...6D} {518, L6}

\bibitem[\protect\citeauthoryear{{van Dishoeck}, {Blake}, {Jansen}  \&
  {Groesbeck}}{{van Dishoeck} et~al.}{1995}]{1995ApJ...447..760V}
{van Dishoeck} E.~F.,  {Blake} G.~A.,  {Jansen} D.~J.,   {Groesbeck} T.~D.,
  1995, \mn@doi [\apj] {10.1086/175915}, \href
  {http://adsabs.harvard.edu/abs/1995ApJ...447..760V} {447, 760}

\bibitem[\protect\citeauthoryear{{van der Tak}, {Schilke}, {M{\"u}ller}, {Lis},
  {Phillips}, {Gerin}  \& {Roueff}}{{van der Tak}
  et~al.}{2002}]{2002A&A...388L..53V}
{van der Tak} F.~F.~S.,  {Schilke} P.,  {M{\"u}ller} H.~S.~P.,  {Lis} D.~C.,
  {Phillips} T.~G.,  {Gerin} M.,   {Roueff} E.,  2002, \mn@doi [\aap]
  {10.1051/0004-6361:20020647}, \href
  {http://adsabs.harvard.edu/abs/2002A%26A...388L..53V} {388, L53}

\makeatother
\end{thebibliography}




\appendix

\section{Spectroscopic data for NHD and ND$_2$}
Table\,\ref{spectro_nhd1}, \ref{spectro_nhd2}, and \ref{spectro_nd2}, derived from the spectroscopic studies of \citet{1997JChPh.107.9289K} , \citet{2013mss..confETH01M}, \citet{1991JChPh..94.3423K} and \citet{2017ApJS..233...15M}, give the Einstein coefficients, transition frequencies and quantum numbers relevant to this work. In the first three columns of each table, the first number is the upper level quantum number and the second one is the lower level quantum number.

\begin{table}
\caption{Frequencies  and Einstein $A_\mathrm{ul}$ coefficients for the NHD ($N_{K_\mathrm{a} K_\mathrm{c}} J: 1_{0 1}\, 3/2 - 0_{0 0}\, 1/2$) hyperfine components.}             
\label{spectro_nhd1}      
\centering                          
\begin{tabular}{ccccc}
\hline \hline
F$_1$ & F$_2$ & F & Frequency & A$_\mathrm{ul}$  \\    
 &  &  & $\mathrm{MHz}$ & s$^{-1}$ \\
\hline
1.5 $-$ 1.5 & 2 $-$ 1 & 3 $-$ 2 & 412618.32 & $6.24 \times 10^{-7}$ \\
1.5 $-$ 1.5 & 2 $-$ 1 & 2 $-$ 1 & 412622.25 & $6.17 \times 10^{-7}$ \\
1.5 $-$ 1.5 & 2 $-$ 1 & 1 $-$ 0 & 412625.37 & $4.61 \times 10^{-7}$ \\
1.5 $-$ 1.5 & 2 $-$ 1 & 1 $-$ 1 & 412627.02 & $3.36 \times 10^{-7}$ \\
0.5 $-$ 1.5 & 0 $-$ 1 & 1 $-$ 0 & 412636.49 & $1.08 \times 10^{-5}$ \\
0.5 $-$ 1.5 & 0 $-$ 1 & 1 $-$ 1 & 412638.14 & $2.14 \times 10^{-5}$ \\
0.5 $-$ 1.5 & 0 $-$ 1 & 1 $-$ 2 & 412641.64 & $1.46 \times 10^{-5}$ \\
1.5 $-$ 1.5 & 1 $-$ 1 & 2 $-$ 1 & 412652.49 & $1.98 \times 10^{-5}$ \\
0.5 $-$ 1.5 & 1 $-$ 2 & 2 $-$ 2 & 412654.43 & $4.39 \times 10^{-7}$ \\
1.5 $-$ 1.5 & 1 $-$ 1 & 2 $-$ 2 & 412655.99 & $3.77 \times 10^{-5}$ \\
0.5 $-$ 1.5 & 1 $-$ 2 & 1 $-$ 1 & 412657.50 & $1.85 \times 10^{-6}$ \\
1.5 $-$ 1.5 & 1 $-$ 1 & 1 $-$ 0 & 412657.92 & $1.77 \times 10^{-5}$ \\
1.5 $-$ 1.5 & 1 $-$ 1 & 1 $-$ 1 & 412659.57 & $4.92 \times 10^{-6}$ \\
1.5 $-$ 1.5 & 1 $-$ 1 & 0 $-$ 1 & 412661.07 & $6.68 \times 10^{-5}$ \\
0.5 $-$ 1.5 & 1 $-$ 2 & 0 $-$ 1 & 412662.22 & $1.18 \times 10^{-5}$ \\
0.5 $-$ 1.5 & 1 $-$ 2 & 2 $-$ 3 & 412662.48 & $2.78 \times 10^{-6}$ \\
0.5 $-$ 1.5 & 1 $-$ 2 & 1 $-$ 2 & 412662.63 & $5.25 \times 10^{-6}$ \\
1.5 $-$ 1.5 & 1 $-$ 1 & 1 $-$ 2 & 412663.07 & $4.33 \times 10^{-5}$ \\
1.5 $-$ 1.5 & 2 $-$ 2 & 3 $-$ 2 & 412672.05 & $2.14 \times 10^{-6}$ \\
1.5 $-$ 1.5 & 2 $-$ 2 & 2 $-$ 1 & 412674.35 & $4.83 \times 10^{-6}$ \\
1.5 $-$ 1.5 & 2 $-$ 2 & 1 $-$ 1 & 412679.12 & $3.57 \times 10^{-5}$ \\
1.5 $-$ 1.5 & 2 $-$ 2 & 2 $-$ 2 & 412679.48 & $3.00 \times 10^{-5}$ \\
1.5 $-$ 1.5 & 2 $-$ 2 & 3 $-$ 3 & 412680.10 & $2.89 \times 10^{-5}$ \\
1.5 $-$ 1.5 & 2 $-$ 2 & 1 $-$ 2 & 412684.26 & $1.18 \times 10^{-5}$ \\
2.5 $-$ 1.5 & 2 $-$ 1 & 3 $-$ 2 & 412685.61 & $8.34 \times 10^{-5}$ \\
1.5 $-$ 1.5 & 2 $-$ 2 & 2 $-$ 3 & 412687.54 & $5.51 \times 10^{-6}$ \\
0.5 $-$ 0.5 & 1 $-$ 1 & 2 $-$ 1 & 412688.06 & $2.95 \times 10^{-6}$ \\
2.5 $-$ 1.5 & 2 $-$ 1 & 2 $-$ 1 & 412690.53 & $5.90 \times 10^{-5}$ \\
0.5 $-$ 0.5 & 1 $-$ 1 & 2 $-$ 2 & 412691.87 & $3.43 \times 10^{-5}$ \\
2.5 $-$ 1.5 & 2 $-$ 1 & 1 $-$ 0 & 412693.93 & $4.65 \times 10^{-5}$ \\
2.5 $-$ 1.5 & 2 $-$ 1 & 2 $-$ 2 & 412694.04 & $3.38 \times 10^{-5}$ \\
2.5 $-$ 1.5 & 2 $-$ 1 & 1 $-$ 1 & 412695.58 & $4.55 \times 10^{-5}$ \\
0.5 $-$ 0.5 & 1 $-$ 1 & 1 $-$ 1 & 412696.26 & $3.72 \times 10^{-5}$ \\
0.5 $-$ 0.5 & 1 $-$ 1 & 1 $-$ 0 & 412697.17 & $3.00 \times 10^{-5}$ \\
2.5 $-$ 1.5 & 2 $-$ 1 & 1 $-$ 2 & 412699.08 & $5.35 \times 10^{-6}$ \\
0.5 $-$ 0.5 & 1 $-$ 1 & 1 $-$ 2 & 412700.08 & $1.46 \times 10^{-5}$ \\
0.5 $-$ 0.5 & 1 $-$ 1 & 0 $-$ 1 & 412700.98 & $9.07 \times 10^{-5}$ \\
0.5 $-$ 0.5 & 1 $-$ 0 & 2 $-$ 1 & 412701.57 & $8.03 \times 10^{-5}$ \\
1.5 $-$ 1.5 & 1 $-$ 2 & 2 $-$ 1 & 412704.60 & $2.01 \times 10^{-7}$ \\
2.5 $-$ 1.5 & 3 $-$ 2 & 4 $-$ 3 & 412708.09 & $1.21 \times 10^{-4}$ \\
2.5 $-$ 1.5 & 3 $-$ 2 & 3 $-$ 2 & 412708.13 & $1.06 \times 10^{-4}$ \\
2.5 $-$ 1.5 & 3 $-$ 2 & 2 $-$ 1 & 412708.92 & $9.93 \times 10^{-5}$ \\
1.5 $-$ 0.5 & 2 $-$ 1 & 3 $-$ 2 & 412709.49 & $8.94 \times 10^{-5}$ \\
1.5 $-$ 1.5 & 1 $-$ 2 & 2 $-$ 2 & 412709.73 & $2.51 \times 10^{-6}$ \\
0.5 $-$ 0.5 & 1 $-$ 0 & 1 $-$ 1 & 412709.77 & $3.17 \times 10^{-5}$ \\
1.5 $-$ 1.5 & 1 $-$ 2 & 1 $-$ 1 & 412711.67 & $4.04 \times 10^{-6}$ \\
1.5 $-$ 0.5 & 2 $-$ 1 & 2 $-$ 1 & 412713.11 & $5.92 \times 10^{-5}$ \\
1.5 $-$ 1.5 & 1 $-$ 2 & 0 $-$ 1 & 412713.17 & $1.35 \times 10^{-5}$ \\
2.5 $-$ 1.5 & 3 $-$ 2 & 2 $-$ 2 & 412714.05 & $2.08 \times 10^{-5}$ \\
0.5 $-$ 0.5 & 1 $-$ 0 & 0 $-$ 1 & 412714.49 & $1.80 \times 10^{-5}$ \\
2.5 $-$ 1.5 & 3 $-$ 2 & 3 $-$ 3 & 412716.18 & $1.48 \times 10^{-5}$ \\
1.5 $-$ 1.5 & 1 $-$ 2 & 1 $-$ 2 & 412716.80 & $1.02 \times 10^{-5}$ \\
1.5 $-$ 0.5 & 2 $-$ 1 & 2 $-$ 2 & 412716.93 & $1.90 \times 10^{-5}$ \\
1.5 $-$ 1.5 & 1 $-$ 2 & 2 $-$ 3 & 412717.78 & $1.36 \times 10^{-5}$ \\
1.5 $-$ 0.5 & 2 $-$ 1 & 1 $-$ 1 & 412717.88 & $3.11 \times 10^{-5}$ \\
1.5 $-$ 0.5 & 2 $-$ 1 & 1 $-$ 0 & 412718.80 & $3.85 \times 10^{-5}$ \\
1.5 $-$ 0.5 & 2 $-$ 1 & 1 $-$ 2 & 412721.70 & $2.17 \times 10^{-6}$ \\
2.5 $-$ 1.5 & 3 $-$ 2 & 2 $-$ 3 & 412722.10 & $6.49 \times 10^{-7}$ \\
1.5 $-$ 0.5 & 2 $-$ 0 & 2 $-$ 1 & 412726.62 & $1.82 \times 10^{-6}$ \\
0.5 $-$ 0.5 & 0 $-$ 1 & 1 $-$ 1 & 412729.00 & $1.88 \times 10^{-5}$ \\
0.5 $-$ 0.5 & 0 $-$ 1 & 1 $-$ 0 & 412729.91 & $9.56 \times 10^{-6}$ \\
\hline
\end{tabular}
\end{table}

\begin{table}
\contcaption{}
\label{cont_spectro_nhd1}      
\centering                          
\begin{tabular}{ccccc}
\hline \hline
F$_1$ & F$_2$ & F & Frequency & A$_\mathrm{ul}$  \\    
 &  &  & $\mathrm{MHz}$ & s$^{-1}$ \\
\hline
1.5 $-$ 0.5 & 2 $-$ 0 & 1 $-$ 1 & 412731.40 & $9.01 \times 10^{-7}$ \\
0.5 $-$ 0.5 & 0 $-$ 1 & 1 $-$ 2 & 412732.81 & $3.30 \times 10^{-5}$ \\
2.5 $-$ 1.5 & 2 $-$ 2 & 3 $-$ 2 & 412739.34 & $5.26 \times 10^{-6}$ \\
0.5 $-$ 0.5 & 0 $-$ 0 & 1 $-$ 1 & 412742.51 & $1.29 \times 10^{-5}$ \\
2.5 $-$ 1.5 & 2 $-$ 2 & 2 $-$ 1 & 412742.64 & $4.44 \times 10^{-6}$ \\
1.5 $-$ 0.5 & 1 $-$ 1 & 2 $-$ 1 & 412743.36 & $1.11 \times 10^{-5}$ \\
2.5 $-$ 0.5 & 3 $-$ 1 & 3 $-$ 2 & 412745.57 & $3.26 \times 10^{-7}$ \\
1.5 $-$ 0.5 & 1 $-$ 1 & 2 $-$ 2 & 412747.17 & $2.05 \times 10^{-5}$ \\
2.5 $-$ 1.5 & 2 $-$ 2 & 3 $-$ 3 & 412747.39 & $2.61 \times 10^{-5}$ \\
2.5 $-$ 0.5 & 3 $-$ 1 & 2 $-$ 1 & 412747.68 & $2.06 \times 10^{-7}$ \\
2.5 $-$ 1.5 & 2 $-$ 2 & 1 $-$ 1 & 412747.68 & $1.18 \times 10^{-5}$ \\
2.5 $-$ 1.5 & 2 $-$ 2 & 2 $-$ 2 & 412747.77 & $1.18 \times 10^{-5}$ \\
1.5 $-$ 0.5 & 1 $-$ 1 & 1 $-$ 0 & 412751.34 & $2.39 \times 10^{-6}$ \\
1.5 $-$ 0.5 & 1 $-$ 1 & 0 $-$ 1 & 412751.93 & $4.06 \times 10^{-6}$ \\
2.5 $-$ 1.5 & 2 $-$ 2 & 1 $-$ 2 & 412752.81 & $6.06 \times 10^{-6}$ \\
1.5 $-$ 0.5 & 1 $-$ 1 & 1 $-$ 2 & 412754.25 & $1.45 \times 10^{-5}$ \\
2.5 $-$ 1.5 & 2 $-$ 2 & 2 $-$ 3 & 412755.82 & $5.98 \times 10^{-6}$ \\
1.5 $-$ 0.5 & 1 $-$ 0 & 2 $-$ 1 & 412756.87 & $1.57 \times 10^{-5}$ \\
1.5 $-$ 0.5 & 1 $-$ 0 & 1 $-$ 1 & 412763.95 & $2.40 \times 10^{-5}$ \\
1.5 $-$ 0.5 & 1 $-$ 0 & 0 $-$ 1 & 412765.45 & $3.68 \times 10^{-5}$ \\
2.5 $-$ 0.5 & 2 $-$ 1 & 3 $-$ 2 & 412776.78 & $6.23 \times 10^{-6}$ \\
2.5 $-$ 0.5 & 2 $-$ 1 & 2 $-$ 1 & 412781.40 & $2.96 \times 10^{-6}$ \\
2.5 $-$ 0.5 & 2 $-$ 1 & 2 $-$ 2 & 412785.21 & $2.78 \times 10^{-6}$ \\
2.5 $-$ 0.5 & 2 $-$ 1 & 1 $-$ 1 & 412786.44 & $2.67 \times 10^{-6}$ \\
2.5 $-$ 0.5 & 2 $-$ 1 & 1 $-$ 0 & 412787.35 & $2.44 \times 10^{-6}$ \\
2.5 $-$ 0.5 & 2 $-$ 1 & 1 $-$ 2 & 412790.25 & $4.40 \times 10^{-7}$ \\
2.5 $-$ 0.5 & 2 $-$ 0 & 2 $-$ 1 & 412794.91 & $3.60 \times 10^{-7}$ \\
\hline
\end{tabular}
\end{table}

\begin{table}
\caption{Frequencies  and Einstein $A_\mathrm{ul}$ coefficients for the NHD ($N_{K_\mathrm{a} K_\mathrm{c}} J: 1_{0 1}\, 1/2 - 0_{0 0}\, 1/2$) hyperfine components.}             
\label{spectro_nhd2}      
\centering                          
\begin{tabular}{ccccc}
\hline \hline
F$_1$ & F$_2$ & F & Frequency & A$_\mathrm{ul}$  \\    
 &  &  & $\mathrm{MHz}$ & s$^{-1}$ \\
\hline
1.5 $-$ 1.5 & 1 $-$ 1 & 0 $-$ 1 & 413473.64 & $3.40 \times 10^{-5}$ \\
1.5 $-$ 1.5 & 1 $-$ 1 & 1 $-$ 0 & 413474.03 & $2.09 \times 10^{-5}$ \\
1.5 $-$ 1.5 & 2 $-$ 1 & 1 $-$ 0 & 413477.63 & $2.15 \times 10^{-6}$ \\
1.5 $-$ 1.5 & 1 $-$ 1 & 1 $-$ 2 & 413479.18 & $1.03 \times 10^{-5}$ \\
1.5 $-$ 1.5 & 2 $-$ 1 & 1 $-$ 1 & 413479.28 & $1.34 \times 10^{-5}$ \\
1.5 $-$ 1.5 & 2 $-$ 1 & 2 $-$ 1 & 413480.36 & $2.80 \times 10^{-5}$ \\
1.5 $-$ 1.5 & 2 $-$ 1 & 1 $-$ 2 & 413482.78 & $7.62 \times 10^{-6}$ \\
1.5 $-$ 1.5 & 1 $-$ 1 & 2 $-$ 1 & 413485.66 & $3.78 \times 10^{-7}$ \\
1.5 $-$ 1.5 & 1 $-$ 1 & 2 $-$ 2 & 413489.16 & $2.22 \times 10^{-5}$ \\
1.5 $-$ 1.5 & 2 $-$ 1 & 3 $-$ 2 & 413493.44 & $3.68 \times 10^{-5}$ \\
1.5 $-$ 1.5 & 1 $-$ 2 & 0 $-$ 1 & 413525.74 & $6.08 \times 10^{-6}$ \\
1.5 $-$ 1.5 & 1 $-$ 2 & 1 $-$ 1 & 413527.78 & $1.11 \times 10^{-6}$ \\
1.5 $-$ 1.5 & 2 $-$ 2 & 1 $-$ 1 & 413531.38 & $4.03 \times 10^{-5}$ \\
1.5 $-$ 1.5 & 2 $-$ 2 & 2 $-$ 1 & 413532.46 & $2.87 \times 10^{-6}$ \\
1.5 $-$ 1.5 & 1 $-$ 2 & 1 $-$ 2 & 413532.91 & $1.57 \times 10^{-5}$ \\
1.5 $-$ 1.5 & 2 $-$ 2 & 1 $-$ 2 & 413536.51 & $7.44 \times 10^{-6}$ \\
1.5 $-$ 1.5 & 2 $-$ 2 & 2 $-$ 2 & 413537.59 & $6.03 \times 10^{-6}$ \\
1.5 $-$ 1.5 & 1 $-$ 2 & 2 $-$ 1 & 413537.76 & $8.30 \times 10^{-6}$ \\
0.5 $-$ 1.5 & 1 $-$ 1 & 1 $-$ 0 & 413538.36 & $1.07 \times 10^{-6}$ \\
0.5 $-$ 1.5 & 1 $-$ 1 & 2 $-$ 1 & 413539.84 & $1.30 \times 10^{-5}$ \\
0.5 $-$ 1.5 & 1 $-$ 1 & 1 $-$ 1 & 413540.01 & $3.46 \times 10^{-7}$ \\
0.5 $-$ 1.5 & 1 $-$ 1 & 0 $-$ 1 & 413542.57 & $1.95 \times 10^{-5}$ \\
1.5 $-$ 1.5 & 1 $-$ 2 & 2 $-$ 2 & 413542.89 & $3.64 \times 10^{-5}$ \\
0.5 $-$ 1.5 & 1 $-$ 1 & 2 $-$ 2 & 413543.34 & $2.66 \times 10^{-5}$ \\
0.5 $-$ 1.5 & 1 $-$ 1 & 1 $-$ 2 & 413543.51 & $2.80 \times 10^{-5}$ \\
0.5 $-$ 1.5 & 0 $-$ 1 & 1 $-$ 0 & 413543.91 & $2.11 \times 10^{-5}$ \\
0.5 $-$ 1.5 & 0 $-$ 1 & 1 $-$ 1 & 413545.56 & $3.48 \times 10^{-5}$ \\
1.5 $-$ 1.5 & 2 $-$ 2 & 2 $-$ 3 & 413545.65 & $3.13 \times 10^{-5}$ \\
1.5 $-$ 1.5 & 2 $-$ 2 & 3 $-$ 2 & 413547.17 & $7.77 \times 10^{-6}$ \\
0.5 $-$ 1.5 & 0 $-$ 1 & 1 $-$ 2 & 413549.06 & $1.15 \times 10^{-5}$ \\
1.5 $-$ 1.5 & 1 $-$ 2 & 2 $-$ 3 & 413550.94 & $4.04 \times 10^{-7}$ \\
1.5 $-$ 1.5 & 2 $-$ 2 & 3 $-$ 3 & 413555.22 & $5.10 \times 10^{-5}$ \\
1.5 $-$ 0.5 & 1 $-$ 1 & 0 $-$ 1 & 413564.50 & $1.61 \times 10^{-5}$ \\
1.5 $-$ 0.5 & 1 $-$ 1 & 1 $-$ 1 & 413566.54 & $5.90 \times 10^{-6}$ \\
1.5 $-$ 0.5 & 1 $-$ 1 & 1 $-$ 0 & 413567.45 & $5.27 \times 10^{-7}$ \\
1.5 $-$ 0.5 & 2 $-$ 1 & 1 $-$ 1 & 413570.14 & $1.32 \times 10^{-5}$ \\
1.5 $-$ 0.5 & 1 $-$ 1 & 1 $-$ 2 & 413570.35 & $2.83 \times 10^{-5}$ \\
1.5 $-$ 0.5 & 2 $-$ 1 & 1 $-$ 0 & 413571.05 & $3.03 \times 10^{-5}$ \\
1.5 $-$ 0.5 & 2 $-$ 1 & 1 $-$ 2 & 413573.95 & $1.60 \times 10^{-6}$ \\
1.5 $-$ 0.5 & 2 $-$ 1 & 2 $-$ 2 & 413575.04 & $4.01 \times 10^{-5}$ \\
1.5 $-$ 0.5 & 1 $-$ 1 & 2 $-$ 1 & 413576.52 & $4.44 \times 10^{-5}$ \\
1.5 $-$ 0.5 & 1 $-$ 0 & 0 $-$ 1 & 413578.02 & $6.42 \times 10^{-5}$ \\
1.5 $-$ 0.5 & 1 $-$ 0 & 1 $-$ 1 & 413580.05 & $3.77 \times 10^{-5}$ \\
1.5 $-$ 0.5 & 1 $-$ 1 & 2 $-$ 2 & 413580.33 & $3.81 \times 10^{-6}$ \\
1.5 $-$ 0.5 & 2 $-$ 0 & 1 $-$ 1 & 413583.65 & $4.45 \times 10^{-6}$ \\
1.5 $-$ 0.5 & 2 $-$ 1 & 3 $-$ 2 & 413584.61 & $2.49 \times 10^{-5}$ \\
1.5 $-$ 0.5 & 2 $-$ 0 & 2 $-$ 1 & 413584.74 & $1.20 \times 10^{-5}$ \\
1.5 $-$ 0.5 & 1 $-$ 0 & 2 $-$ 1 & 413590.03 & $4.53 \times 10^{-6}$ \\
0.5 $-$ 1.5 & 1 $-$ 2 & 2 $-$ 1 & 413591.95 & $1.05 \times 10^{-6}$ \\
0.5 $-$ 1.5 & 1 $-$ 2 & 1 $-$ 1 & 413592.11 & $2.26 \times 10^{-5}$ \\
0.5 $-$ 1.5 & 1 $-$ 2 & 0 $-$ 1 & 413594.68 & $8.92 \times 10^{-5}$ \\
0.5 $-$ 1.5 & 1 $-$ 2 & 2 $-$ 2 & 413597.08 & $1.28 \times 10^{-5}$ \\
0.5 $-$ 1.5 & 1 $-$ 2 & 1 $-$ 2 & 413597.24 & $5.58 \times 10^{-5}$ \\
0.5 $-$ 1.5 & 0 $-$ 2 & 1 $-$ 1 & 413597.66 & $3.38 \times 10^{-6}$ \\
0.5 $-$ 1.5 & 0 $-$ 2 & 1 $-$ 2 & 413602.79 & $8.33 \times 10^{-6}$ \\
0.5 $-$ 1.5 & 1 $-$ 2 & 2 $-$ 3 & 413605.13 & $6.04 \times 10^{-5}$ \\
0.5 $-$ 0.5 & 1 $-$ 1 & 1 $-$ 1 & 413630.87 & $5.55 \times 10^{-6}$ \\
0.5 $-$ 0.5 & 1 $-$ 1 & 1 $-$ 0 & 413631.78 & $5.20 \times 10^{-6}$ \\
0.5 $-$ 0.5 & 1 $-$ 1 & 0 $-$ 1 & 413633.43 & $1.01 \times 10^{-5}$ \\
\hline
\end{tabular}
\end{table}

\begin{table}
\contcaption{}
\label{cont_spectro_nhd2}      
\centering                          
\begin{tabular}{ccccc}
\hline \hline
F$_1$ & F$_2$ & F & Frequency & A$_\mathrm{ul}$  \\    
 &  &  & $\mathrm{MHz}$ & s$^{-1}$ \\
\hline
0.5 $-$ 0.5 & 1 $-$ 1 & 2 $-$ 2 & 413634.52 & $3.32 \times 10^{-7}$ \\
0.5 $-$ 0.5 & 1 $-$ 1 & 1 $-$ 2 & 413634.68 & $1.56 \times 10^{-6}$ \\
0.5 $-$ 0.5 & 0 $-$ 1 & 1 $-$ 1 & 413636.42 & $6.42 \times 10^{-6}$ \\
0.5 $-$ 0.5 & 0 $-$ 1 & 1 $-$ 0 & 413637.33 & $1.93 \times 10^{-6}$ \\
0.5 $-$ 0.5 & 0 $-$ 1 & 1 $-$ 2 & 413640.24 & $2.46 \times 10^{-5}$ \\
0.5 $-$ 0.5 & 1 $-$ 0 & 2 $-$ 1 & 413644.22 & $6.26 \times 10^{-6}$ \\
0.5 $-$ 0.5 & 1 $-$ 0 & 1 $-$ 1 & 413644.38 & $3.55 \times 10^{-7}$ \\
0.5 $-$ 0.5 & 1 $-$ 0 & 0 $-$ 1 & 413646.95 & $1.66 \times 10^{-6}$ \\
0.5 $-$ 0.5 & 0 $-$ 0 & 1 $-$ 1 & 413649.94 & $8.48 \times 10^{-6}$ \\
\hline
\end{tabular}
\end{table}

\begin{table}
\caption{Frequencies  and Einstein $A_\mathrm{ul}$ coefficients for the para-ND$_2$ ($N_{K_\mathrm{a} K_\mathrm{c}} J: 1_{1 1}\, 1/2 - 0_{0 0}\, 1/2$) hyperfine components.}             
\label{spectro_nd2}      
\centering                          
\begin{tabular}{c c  c c}        
\hline\hline                 
F$_1$ & F & Frequency  & A$_\mathrm{ul}$  \\   
  &  & $\mathrm{MHz}$ & s$^{-1}$ \\
\hline 
1.5 $-$ 1.5 & 0.5 $-$ 0.5 & 530980.77 & $4.63 \times 10^{-4}$ \\
1.5 $-$ 1.5 & 1.5 $-$ 0.5 & 530981.74 & $2.95 \times 10^{-4}$ \\
1.5 $-$ 1.5 & 0.5 $-$ 1.5 & 530985.03 & $4.17 \times 10^{-4}$ \\
1.5 $-$ 1.5 & 1.5 $-$ 1.5 & 530986.00 & $3.95 \times 10^{-4}$ \\
1.5 $-$ 1.5 & 2.5 $-$ 1.5 & 530987.69 & $3.19 \times 10^{-4}$ \\
1.5 $-$ 1.5 & 1.5 $-$ 2.5 & 530995.83 & $2.89 \times 10^{-4}$ \\
1.5 $-$ 1.5 & 2.5 $-$ 2.5 & 530997.52 & $9.00 \times 10^{-4}$ \\
1.5 $-$ 0.5 & 0.5 $-$ 1.5 & 531031.41 & $1.82 \times 10^{-4}$ \\
1.5 $-$ 0.5 & 1.5 $-$ 1.5 & 531032.38 & $5.44 \times 10^{-4}$ \\
1.5 $-$ 0.5 & 2.5 $-$ 1.5 & 531034.07 & $7.05 \times 10^{-4}$ \\
1.5 $-$ 0.5 & 0.5 $-$ 0.5 & 531035.67 & $8.63 \times 10^{-4}$ \\
1.5 $-$ 0.5 & 1.5 $-$ 0.5 & 531036.64 & $4.01 \times 10^{-4}$ \\
0.5 $-$ 1.5 & 1.5 $-$ 0.5 & 531040.13 & $6.72 \times 10^{-5}$ \\
0.5 $-$ 1.5 & 0.5 $-$ 0.5 & 531041.17 & $7.54 \times 10^{-4}$ \\
0.5 $-$ 1.5 & 1.5 $-$ 1.5 & 531044.39 & $4.67 \times 10^{-4}$ \\
0.5 $-$ 1.5 & 0.5 $-$ 1.5 & 531045.43 & $7.74 \times 10^{-4}$ \\
0.5 $-$ 1.5 & 1.5 $-$ 2.5 & 531054.22 & $1.25 \times 10^{-3}$ \\
0.5 $-$ 0.5 & 1.5 $-$ 1.5 & 531090.77 & $4.71 \times 10^{-5}$ \\
0.5 $-$ 0.5 & 0.5 $-$ 1.5 & 531091.81 & $3.38 \times 10^{-4}$ \\
0.5 $-$ 0.5 & 1.5 $-$ 0.5 & 531095.03 & $9.12 \times 10^{-5}$ \\
0.5 $-$ 0.5 & 0.5 $-$ 0.5 & 531096.06 & $5.90 \times 10^{-5}$ \\
\hline                        
\end{tabular}
\end{table}


\bsp	
\label{lastpage}
\end{document}